\documentclass[openacc]{rsproca_new}

\usepackage{units}
\usepackage[numbers]{natbib}
\UseRawInputEncoding

\jname{rspa}
\Journal{Proc R Soc A\ }

\begin{document}

\title{Realization of Bullard's disc dynamo}

\author{Raúl Alejandro Avalos-Zúñiga$^{1}$, \\
J\={a}nis Priede$^{2,3}$}

\address{$^{1}$Research Centre in Applied Science and Advanced technology,
Instituto Politécnico Nacional (CICATA - Querétaro),
Cerro Blanco 141, Colinas del Cimatario, Querétaro, Mexico\\
$^{2}$Fluid and Complex Systems Research Centre, Coventry University, \\
Coventry, CV1 5FB, United Kingdom\\
$^{3}$Department of Physics, University of Latvia, Riga, LV-1004, Latvia}

\subject{electromagnetism}


\corres{J. Priede\\
\email{j.priede@coventry.ac.uk}}


\begin{abstract}
We report experimental results from three successful runs of a Bullard-type
homopolar disc dynamo. The set-up consisted of a copper disc with
a radius of $\unit[30]{cm}$ and thickness of $\unit[3]{cm}$ which
was placed co-axially beneath a flat, multi-arm spiral coil of the
same size and connected to it electrically at the centre and along
the circumference by sliding liquid-metal contacts. The magnetic field
was measured using Hall probes which were fixed on the top face of
the coil. We measured also the radial voltage drop across the coil.
When the disc rotation rate reached $\Omega\approx\unit[7]{Hz,}$
the magnetic field increased steeply approaching $B_{0}\approx\unit[40]{mT}$
in the central part of the coil. This field was more than two orders
of magnitude stronger than the background magnetic field. In the first
two runs, the electromagnetic torque braking the disc in the dynamo
regime exceeded the breakdown torque of the electric motor driving
the disc. As a result, the motor stalled and the dynamo was interrupted.
Stalling did not occur in the third run when the driving frequency
was set higher and increased faster. We also propose an extended disc
dynamo model which qualitatively reproduces the experimental results. 
\end{abstract}

\maketitle

\section{Introduction}

Bullard's disc dynamo \citep{Bullard1955} is arguably the simplest
model of the magnetohydrodynamic dynamo. It is often used to illustrate
the self-excitation of the magnetic field by moving conductors \citep{Moffatt1978}.
This is how the magnetic fields of the Earth, the Sun and many other
cosmic bodies are thought to come about \citep{Beck1996}. In its
basic form, the Bullard dynamo consists of a solid metal disc and
a wire: the former spinning about its central axis and the latter
twisted around and connected through sliding contacts to the rim and
axis of the disc. If the disc spins sufficiently fast and in the right
direction, such a set-up can amplify the electric current circulating
in the system and, thus, the associated magnetic field. This happens
when the rotation rate of the disc exceeds a certain critical threshold
above which the potential difference induced across the disc exceeds
the voltage drop caused by the ohmic resistance of the system. Then
the current starts to grow exponentially in time resulting in the
self-excitation of the magnetic field. The growth stops when the braking
electromagnetic torque becomes so strong that it slows down the disc.
This is how the dynamo would operate in the ideal case with no background
magnetic field.

In the presence of a background magnetic field, the dynamo may manifest
itself somewhat differently. In this case, a current is induced in
the set-up as soon as the disc starts to rotate. If the disc rotates
in the right direction, the induced current amplifies the background
magnetic field which, in turn, amplifies the current. Thus, the rate
of amplification increases with the speed of rotation and becomes
formally infinite at the dynamo threshold. 

Despite its simplicity, the implementation of the disc dynamo is faced
with severe technical challenges. The main problem is the sliding
electrical contacts which are required to convey the current between
the rim and the axis of the rotating disc. The electrical resistance
of sliding contacts, which are usually made of solid graphite brushes,
is typically several orders of magnitude higher than that of the rest
of the set-up. This results in unrealistically high rotation rates
which are required for the dynamo to operate. Therefore, in contrast
to the fluid dynamos, which have been realised in several laboratory
experiments using liquid metal \citep{Gailitis2001,Stieglitz2001,Monchaux2007}
(\citep{Stefani2019} in preparation), the disc dynamo was thought
to be technically unfeasible \citep{Raedler2002,Dormy2007,Lorrain2007}.

We overcome this problem by using sliding liquid-metal electrical
contacts which are similar to those employed previously in the homopolar
motors and generators \citep{Inall1968,Maribo1987,Maribo2010} as
well as in the laboratory model of Herzenberg dynamo \citep{Herzenberg1958}
built by Lowes and Wilkinson \citep{Lowes1963,Lowes1968}. The set-up
consists of a coil made of a stationary copper disc which is divided
into spiral-shaped sections by thin slits \citep{Priede2013}. The
coil is placed co-axially above the solid copper disc and connected
to the latter by sliding liquid-metal contacts. The slits make the
conductivity of the coil anisotropic which allows this essentially
axially symmetric dynamo to generate an axially symmetric magnetic
field \citep{Plunian2020,Alboussiere2022}. 

In this paper, which is organised as follows, we report results from
three successful runs of such a dynamo. Set-up is described in Sec.
\ref{sec:setup}. Experimental results are presented in Sec. \ref{sec:results}.
In Sec. \ref{sec:model}, we introduce an extended disc dynamo model
which is used in Sec. \ref{sec:discussion} for the interpretation
and analysis of experimental results. Summary and conclusions are
presented in Sec. \ref{sec:conclusion}.

\section{\label{sec:setup}Experiment set-up}

\begin{figure}
\noindent \centering{}\includegraphics[width=0.35\columnwidth]{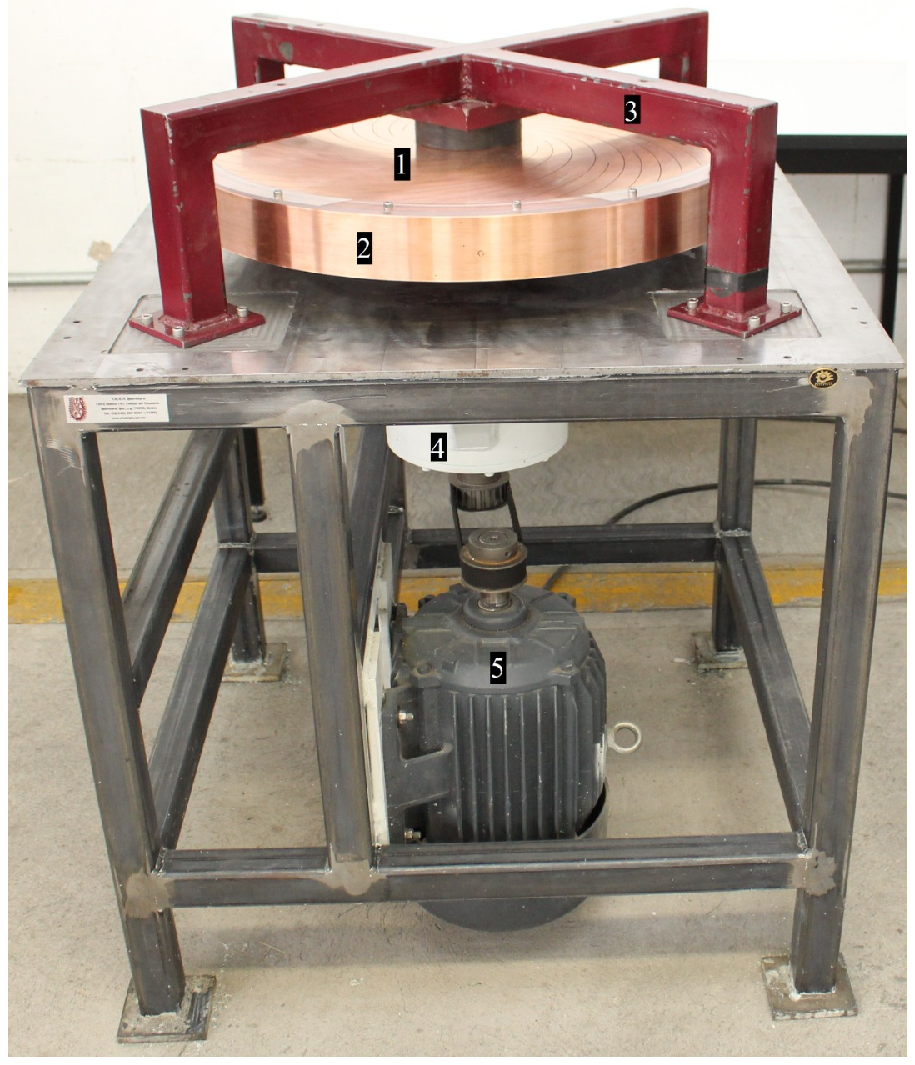}\put(5,5){(a)}\qquad{}\includegraphics[bb=14bp -200bp 545bp 299bp,clip,width=0.45\columnwidth]{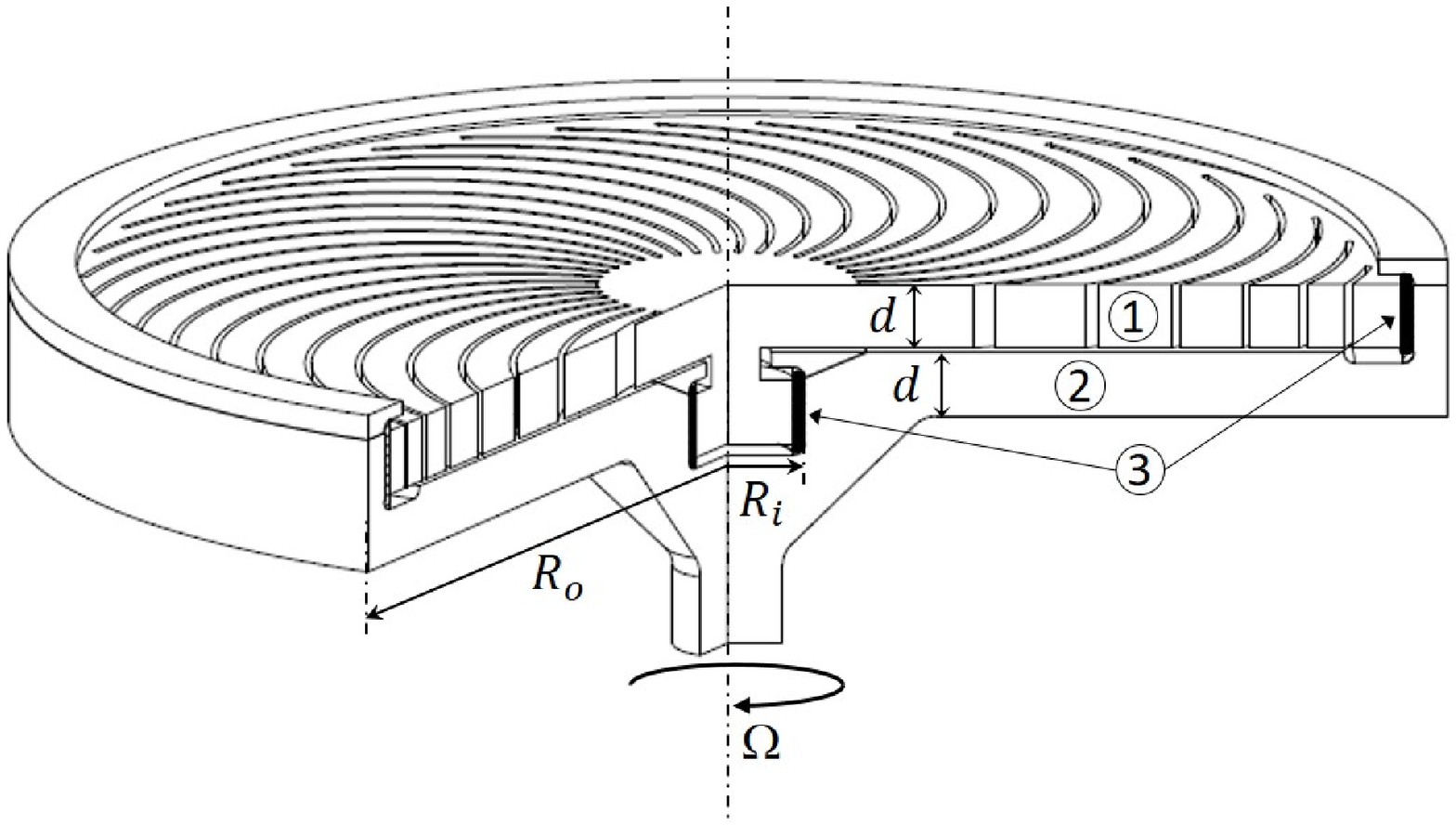}\put(-20,100){(b)}\caption{\label{fig:schm}(a) Set-up consisting of coil (1), disc (2), iron
frame (3), gearing system (4), and AC motor (5); (b) a cross-sectional
view showing coil (1), disc (2), and sliding liquid-metal contacts
(3).}
\end{figure}

The dynamo set-up shown in Fig. \ref{fig:schm} consists of a rotating
copper disc and a coil. The latter is made of a flat copper cylinder
of radius $\unit[r_{o}=30]{cm}$ and thickness $\unit[d=3]{cm}$ which
is sectioned starting from the radius $\tilde{r}_{i}=\unit[7.5]{cm}$
by $40$ logarithmic spiral slits with a constant pitch angle $\alpha\approx58^{\circ}.$
The disc has an annular channel along the rim and a cylindrical cavity
at the centre. The coil, which has a cylindrical solid electrode protruding
$\unit[4]{cm}$ out from the centre of its bottom face, is placed
$\unit[3]{mm}$ above the disc. At the inner and outer radii, $r_{i}=\unit[4.5]{cm}$
and $\unit[r_{o}=30]{cm,}$ there are two annular gaps of width $\unit[\delta=0.25]{mm}$
and height $\unit[d=3]{cm}$ which separate the coil from the disc.
These gaps were filled with the eutectic alloy of $\mathit{GaInSn,}$
which is liquid at room temperature. The coil and disc are held by
electrically insulated iron supports. Note that, in the initial design
with $\unit[3]{mm}$ annular gaps, electrical contacts failed before
the dynamo threshold was attained \citep{Avalos-Zuniga2017}. 

The disc was driven by a $\unit[3]{HP}$ $(\unit[2.2]{kW)}$ 6-pole
AC motor which had a synchronous rotation rate of $\unit[1200]{RPM}$
at $\unit[60]{Hz}$ input frequency. The sense of rotation was opposite
to the orientation of the spiral arms of the coil which corresponds
to a clock-wise rotation in the set-up shown in Fig. \ref{fig:schm}(b).
The rotation rate was changed using a variable frequency drive (VFD)
Delta VFD022EL23A with a rated output current of $\unit[11]{A}$.
The VFD output frequency was controlled from a PC using the LabVIEW
software. With default settings, the output voltage was reduced directly
with the driving frequency. In this standard VFD regime, the driving
torque was maintained constant while the motor power dropped off directly
with the rotation rate. It means that the maximal power the motor
can produce, when operated sufficiently close to the expected dynamo
threshold of $\unit[600]{RPM}$ $(\unit[10]{Hz),}$ is only half of
its rated power, i.e., $\unit[\approx1]{kW.}$ The actual output power
may be significantly lower when the rotation rate starts to drop because
of too high a load.

The rotation rate was determined using a self-made opto-mechanical
tachometer which consisted of a disc with periodic slits with a light
source on one side and a photodiode on the other. The magnetic field
was measured using the THM1176 3-axis Hall Magnetometer of Metrolab
with the low- and medium-field probes THM1176-LF and THM1176-MF. These
probes have an active volume and the upper magnetic field strength
of $\unit[6\text{\ensuremath{\times}}3.4\text{\ensuremath{\times}}3]{mm^{3},}$$\unit[8]{mT}$
and $\unit[16.5\text{\ensuremath{\times}}5\text{\ensuremath{\times}}2.3]{mm}^{3},$
$\unit[100]{mT,}$ respectively. The probes were fixed with adhesive
tape on the upper face of the coil next to one of the crossed iron
arms holding the coil. Three components of field were acquired for
a period of several minutes at the rate of $30$ samples per second.
The voltage between the inner and outer radii of the spiral arms was
measured \textbf{} using a digital multimeter Keithley 2100.

\section{\label{sec:results}Experimental results}

\begin{figure}
\begin{centering}
\includegraphics[width=0.33\columnwidth]{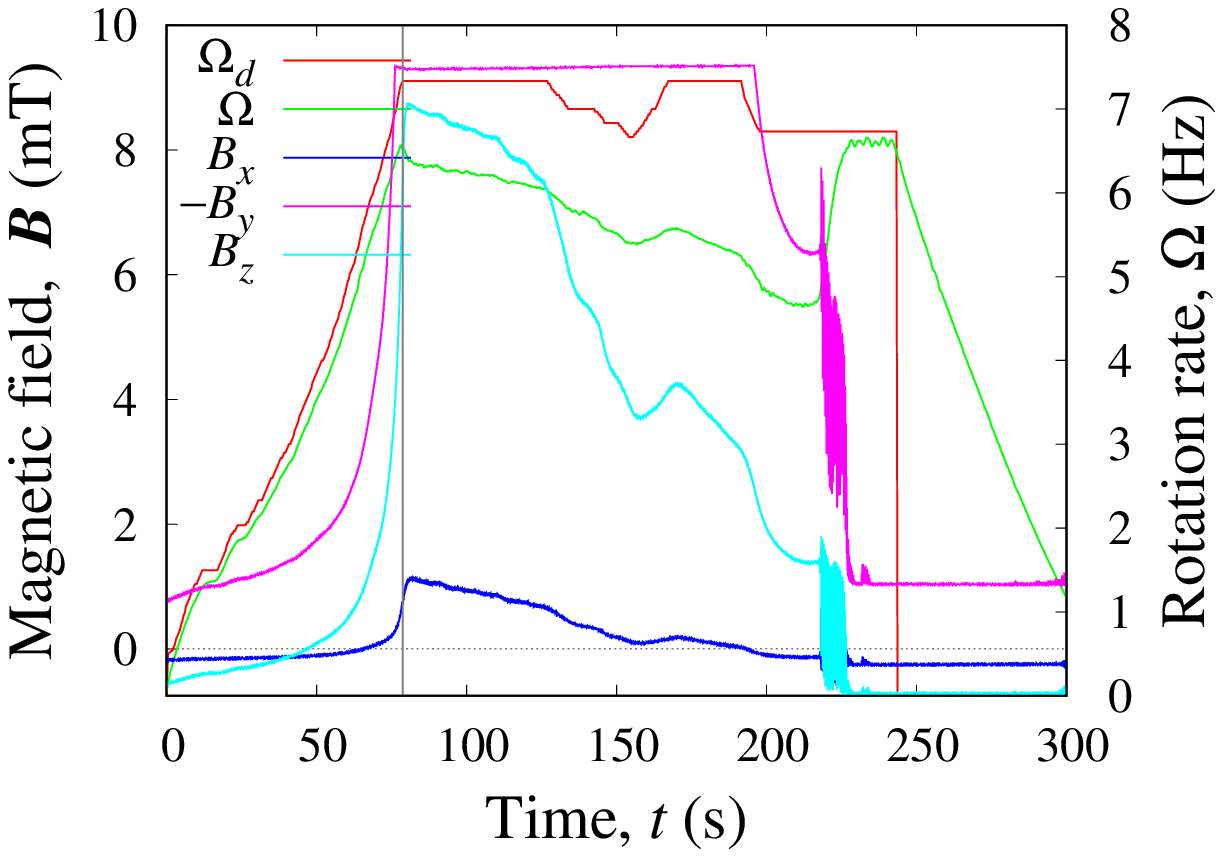}\put(-125,5){(a)}\includegraphics[width=0.33\columnwidth]{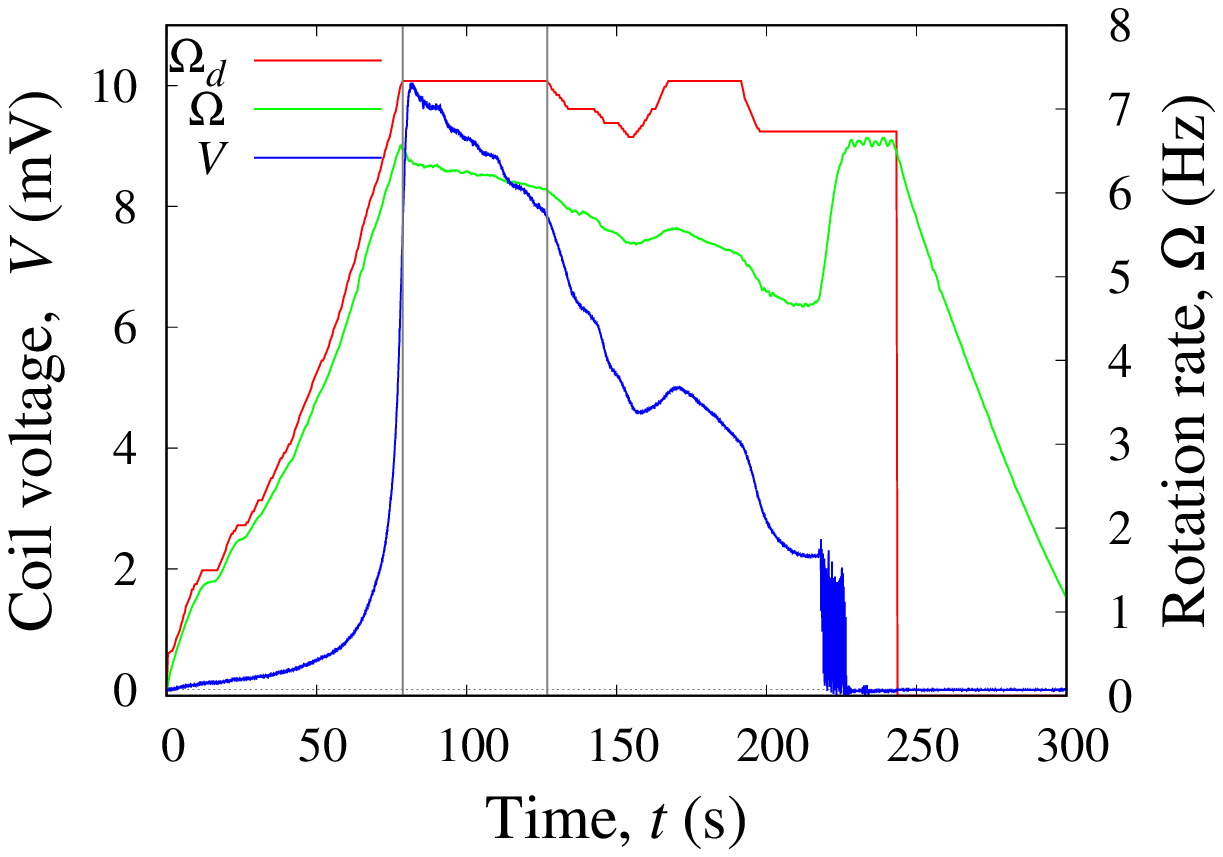}\put(-125,5){(b)}\includegraphics[width=0.33\columnwidth]{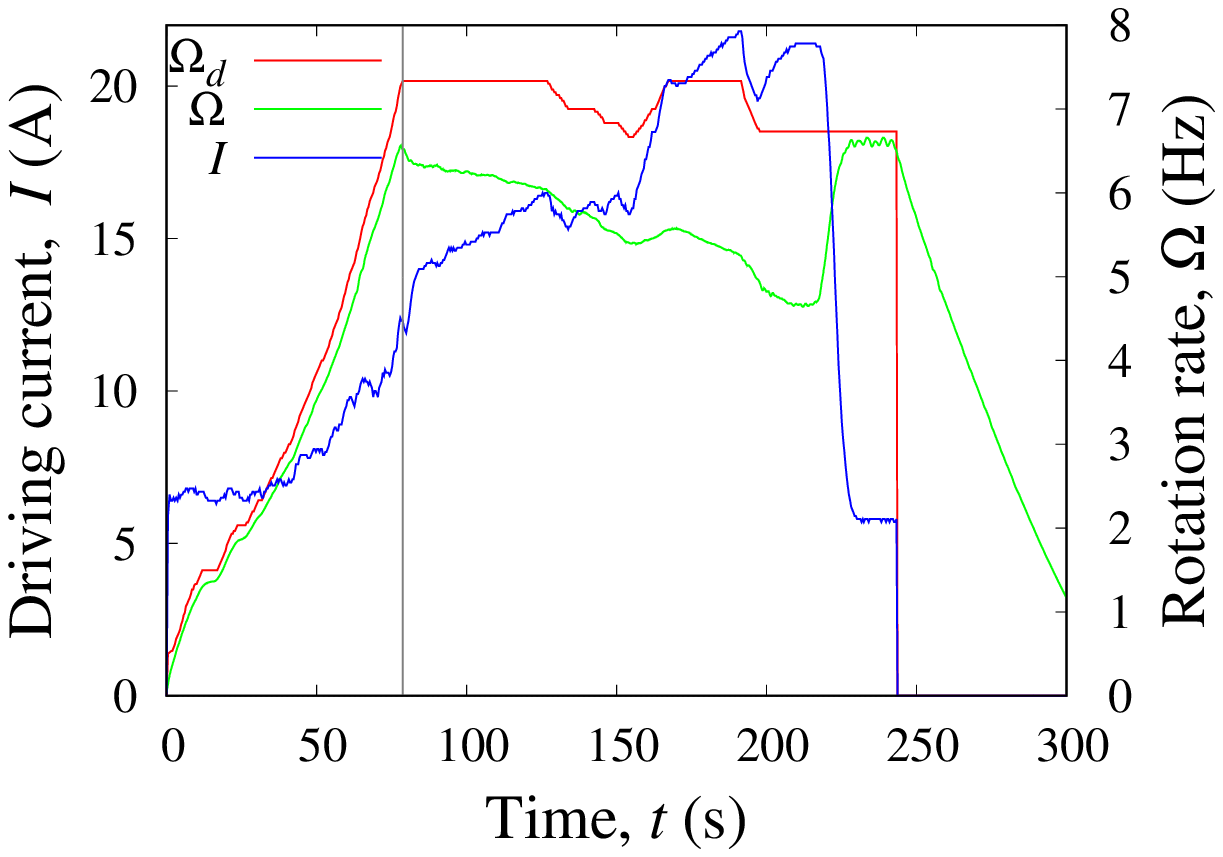}\put(-125,5){(c)}
\par\end{centering}
\begin{centering}
\includegraphics[width=0.33\columnwidth]{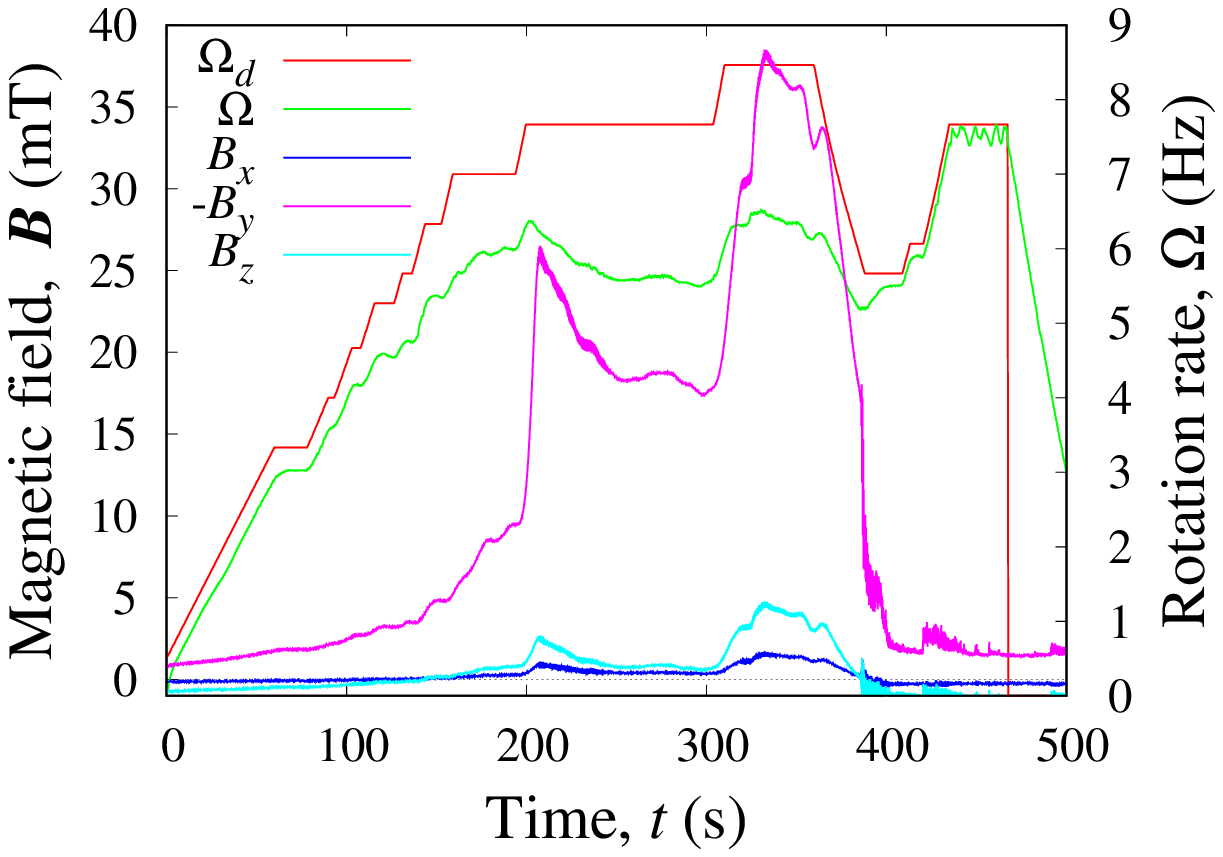}\put(-125,5){(d)}\includegraphics[width=0.33\columnwidth]{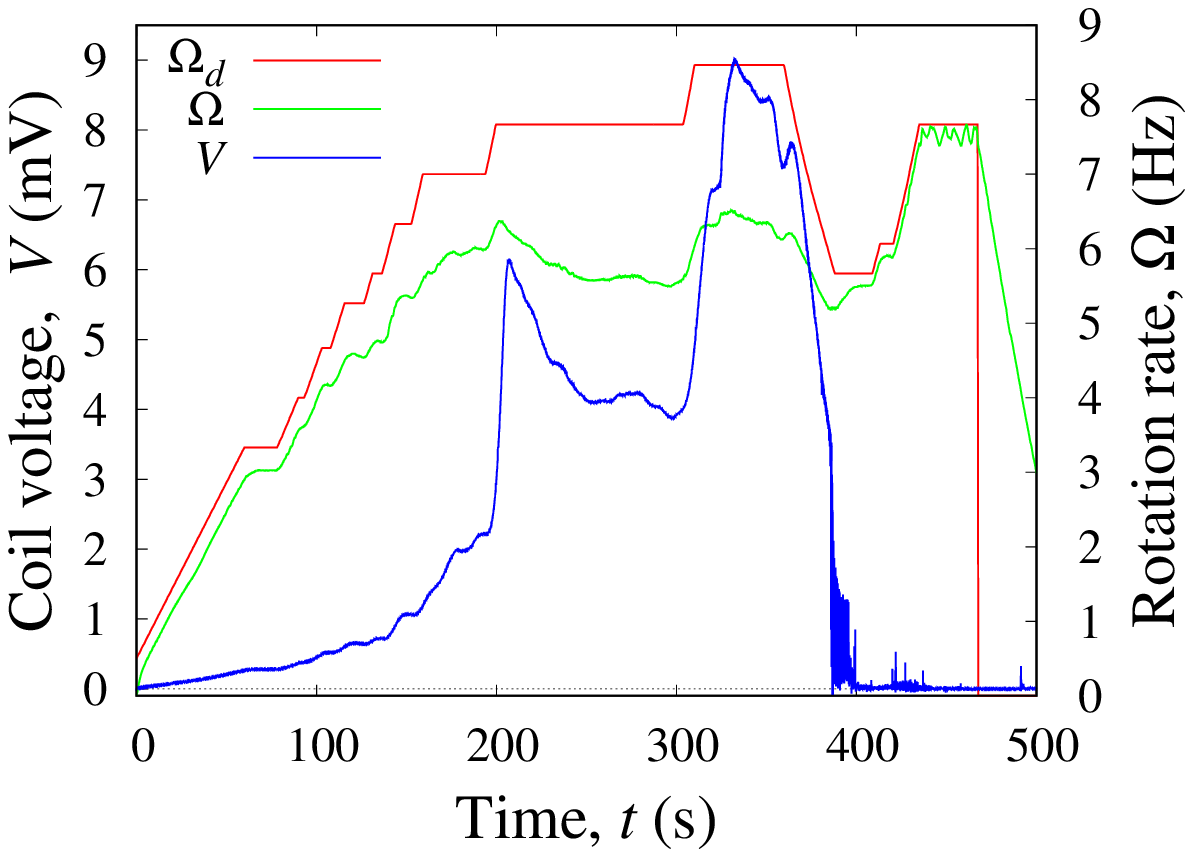}\put(-125,5){(e)}\includegraphics[width=0.33\columnwidth]{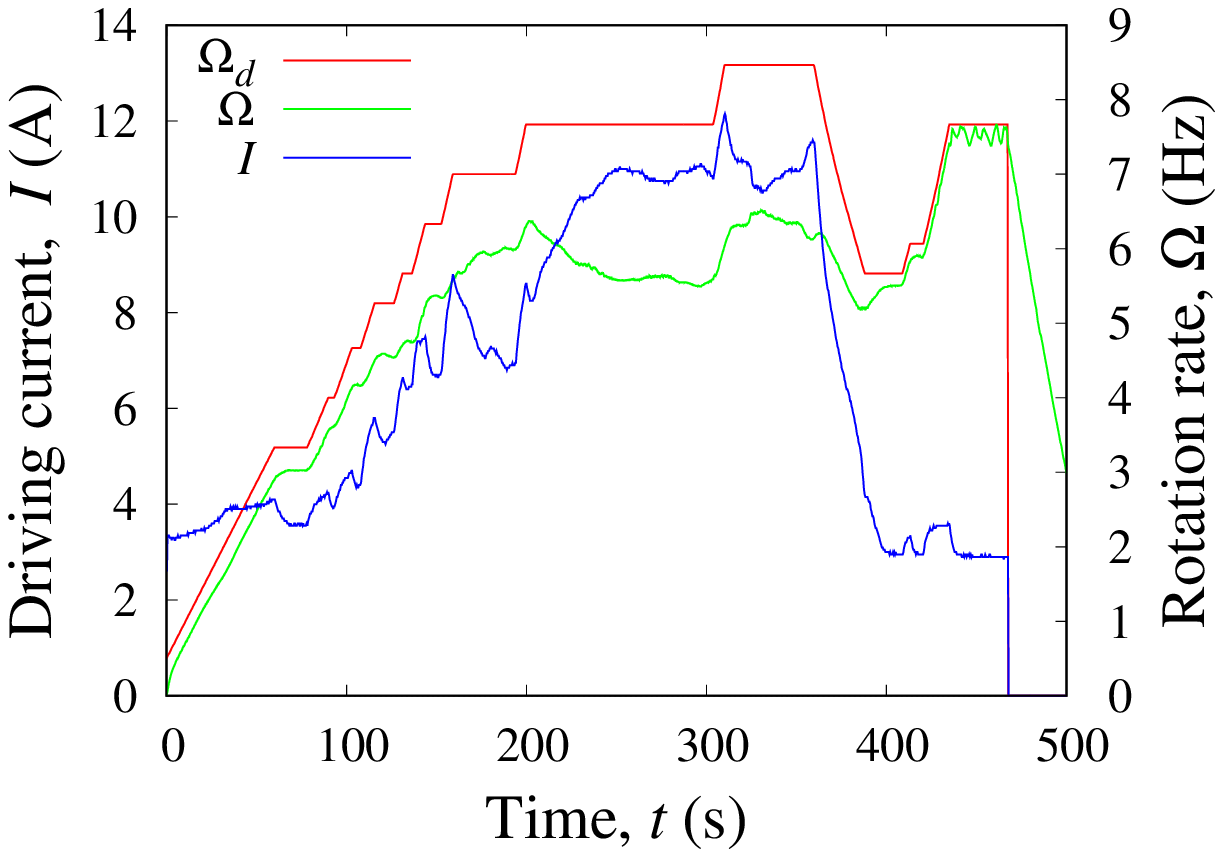}\put(-125,5){(f)}
\par\end{centering}
\begin{centering}
\includegraphics[width=0.33\columnwidth]{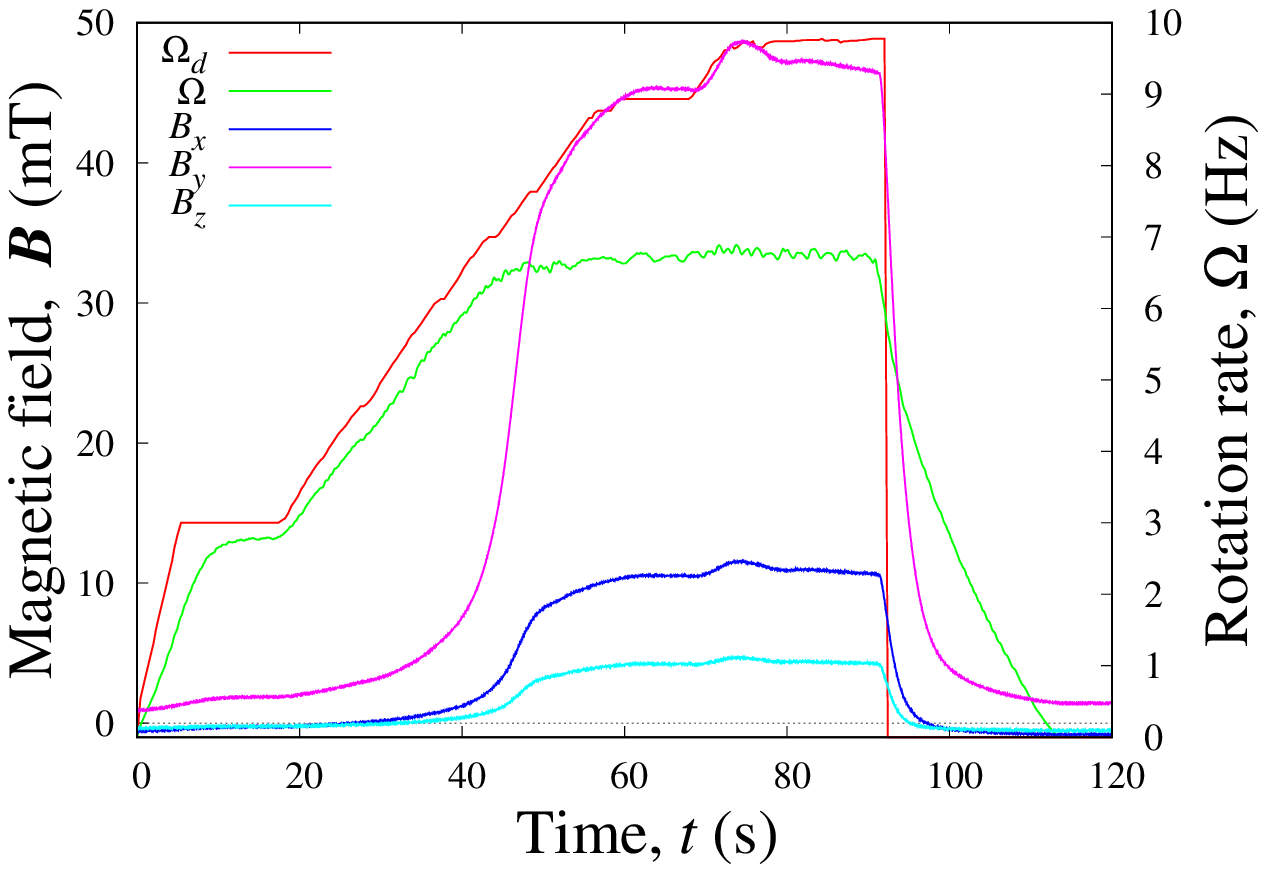}\put(-125,5){(g)}\includegraphics[width=0.33\columnwidth]{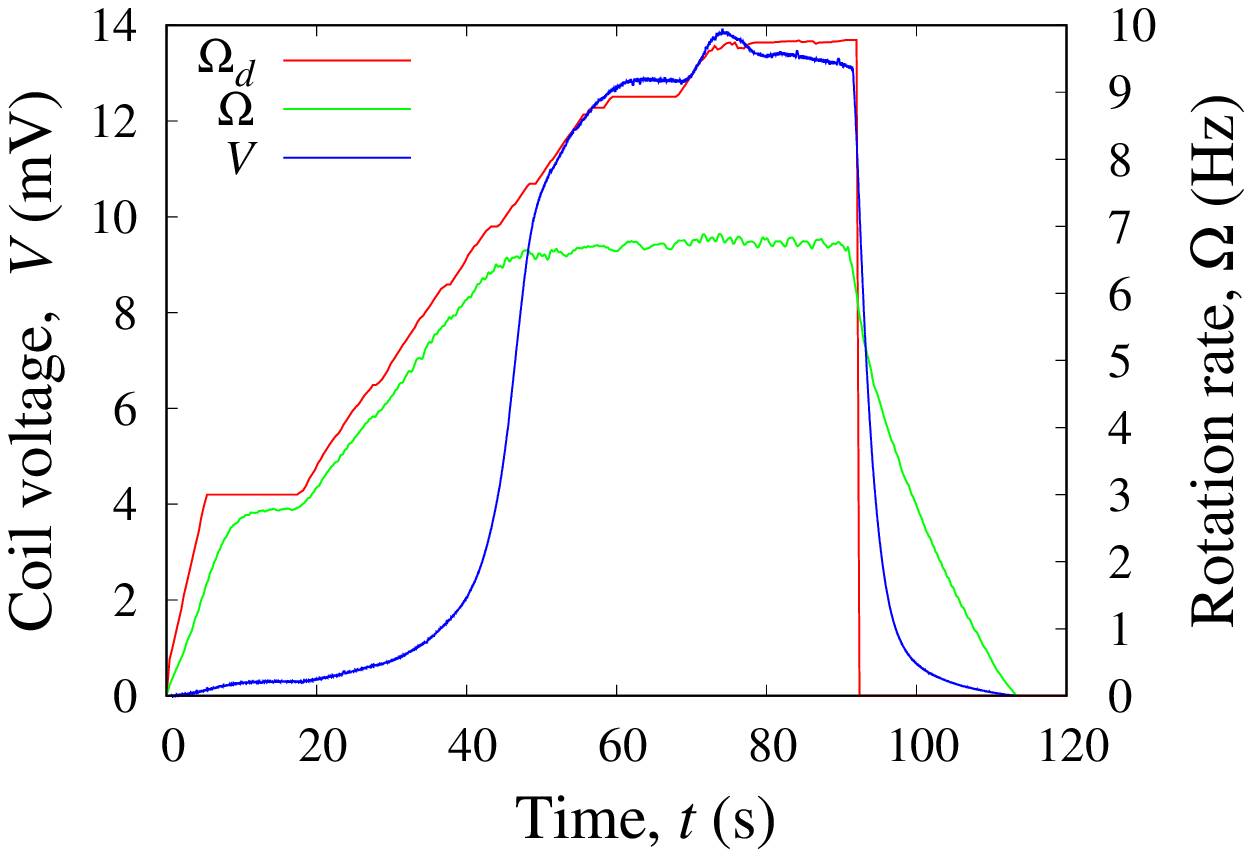}\put(-125,5){(h)}\includegraphics[width=0.33\columnwidth]{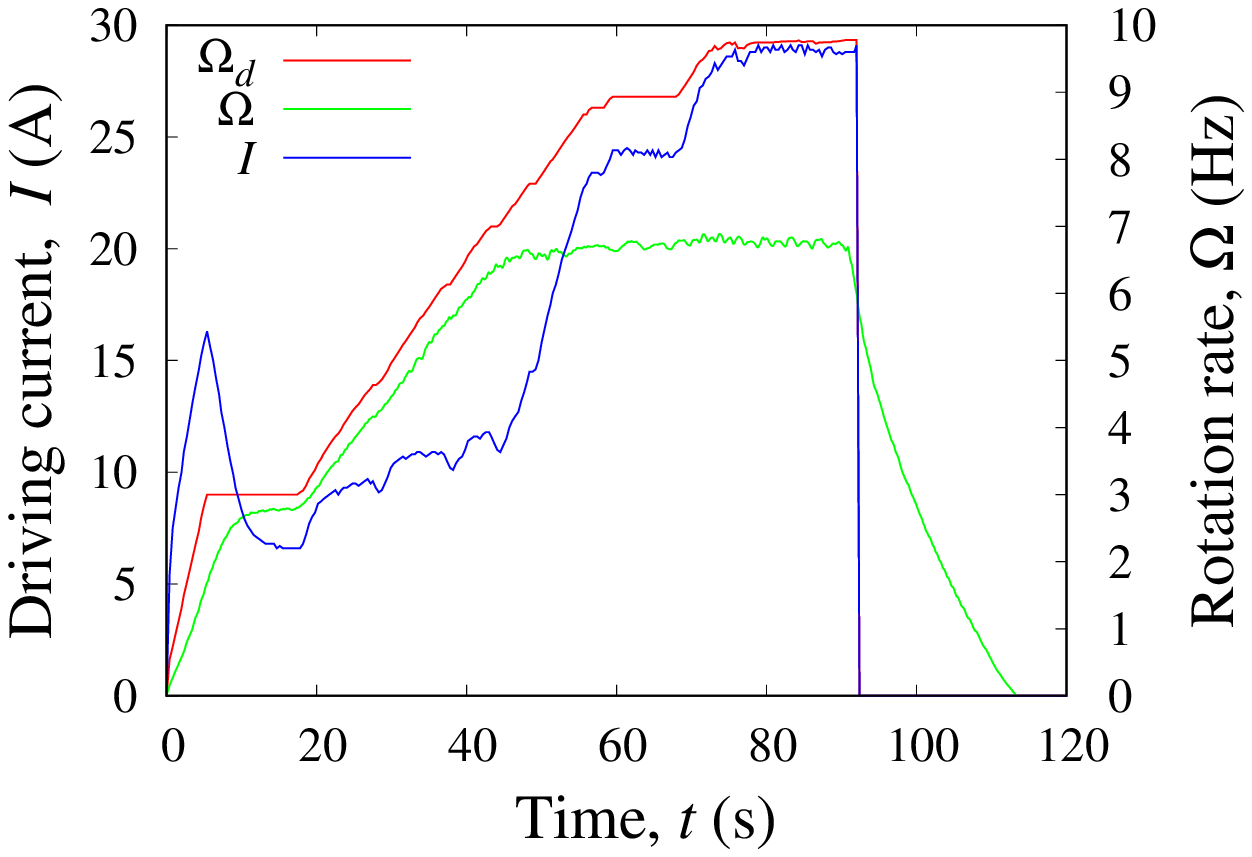}\put(-125,5){(i)}
\par\end{centering}
\caption{\label{fig:tim-ser}The magnetic field $(B_{x},B_{y},B_{z})$ at the
inner radius of the coil (a,d,g), the voltage between the centre and
the rim of the coil (b,e,h), and the electric current supplied to
the motor (c,f,i) along with the driving $(\Omega_{d})$ and rotation
$(\Omega)$ frequencies recorded during the first (a,b,c), second
(d,e,f) and third (g,h,i) runs.}
\end{figure}

In this section, we present experimental results for three runs in
which the disc rotation rate was measured together with the induced
voltage and the magnetic field at the upper face of the coil. The
runs differ mainly by the way in which the frequency driving the motor
was varied. In the first run, for which the results are shown in the
top row of Fig. \ref{fig:tim-ser}, the driving frequency was ramped
up from zero to $\Omega_{d}=\unit[7.33]{Hz}$ nearly linearly in $\approx\unit[79]{s}.$
Although the rotation rate $\Omega$ closely followed $\Omega_{d},$
there was a difference between both frequencies. This difference increased
as the magnetic field became stronger. In this run, the magnetic field
was measured using only the low-field probe which was placed in the
vicinity of the inner radius of the spiral slits. Before the disc
started to rotate, a background field $\vec{B}_{0}\approx\unit[-(0.18,0.77,0.57)]{mT}$
with $y$ component directed downwards and $x$ and $z$ components
in the plane of the coil was detected. This field, which was more
than an order of magnitude stronger than Earth's magnetic field, was
obviously due to the iron frame holding the coil. 

\begin{figure}
\begin{centering}
\includegraphics[width=0.5\columnwidth]{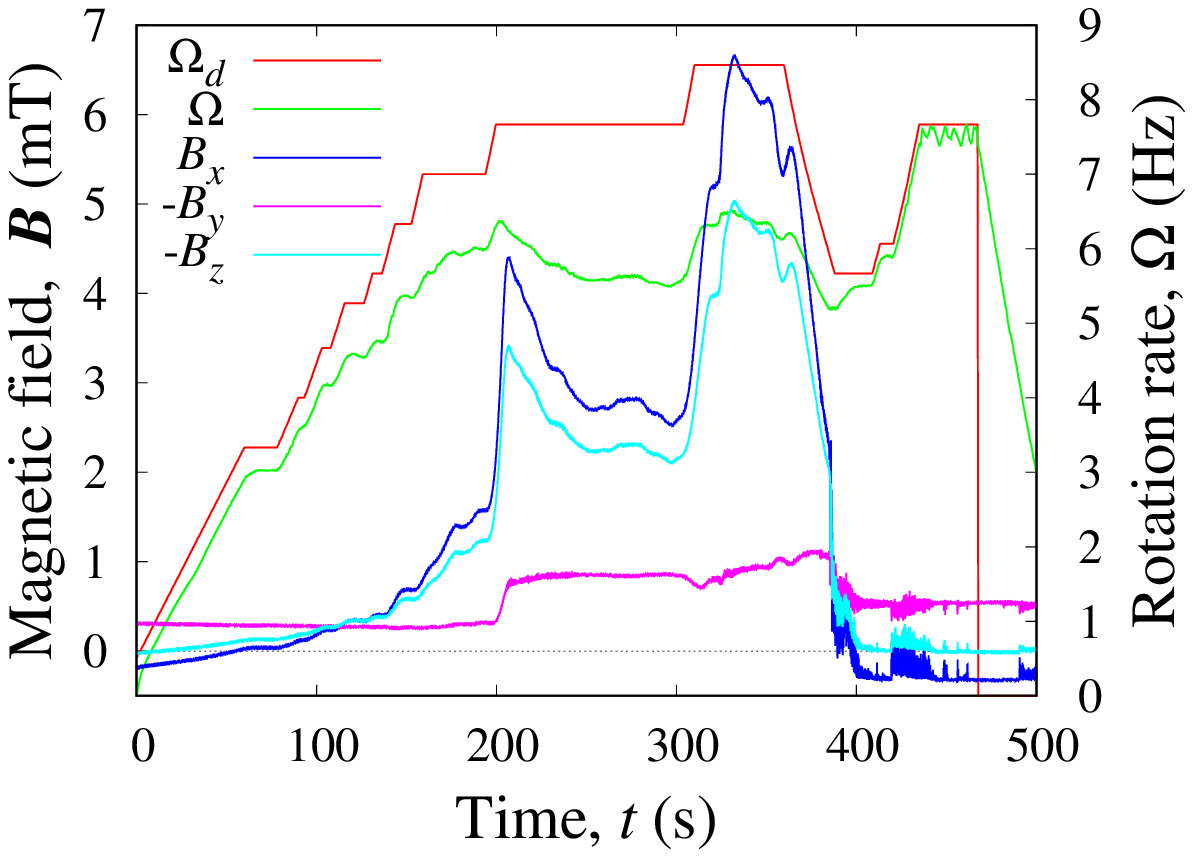}\put(-185,5){(a)}\includegraphics[width=0.5\columnwidth]{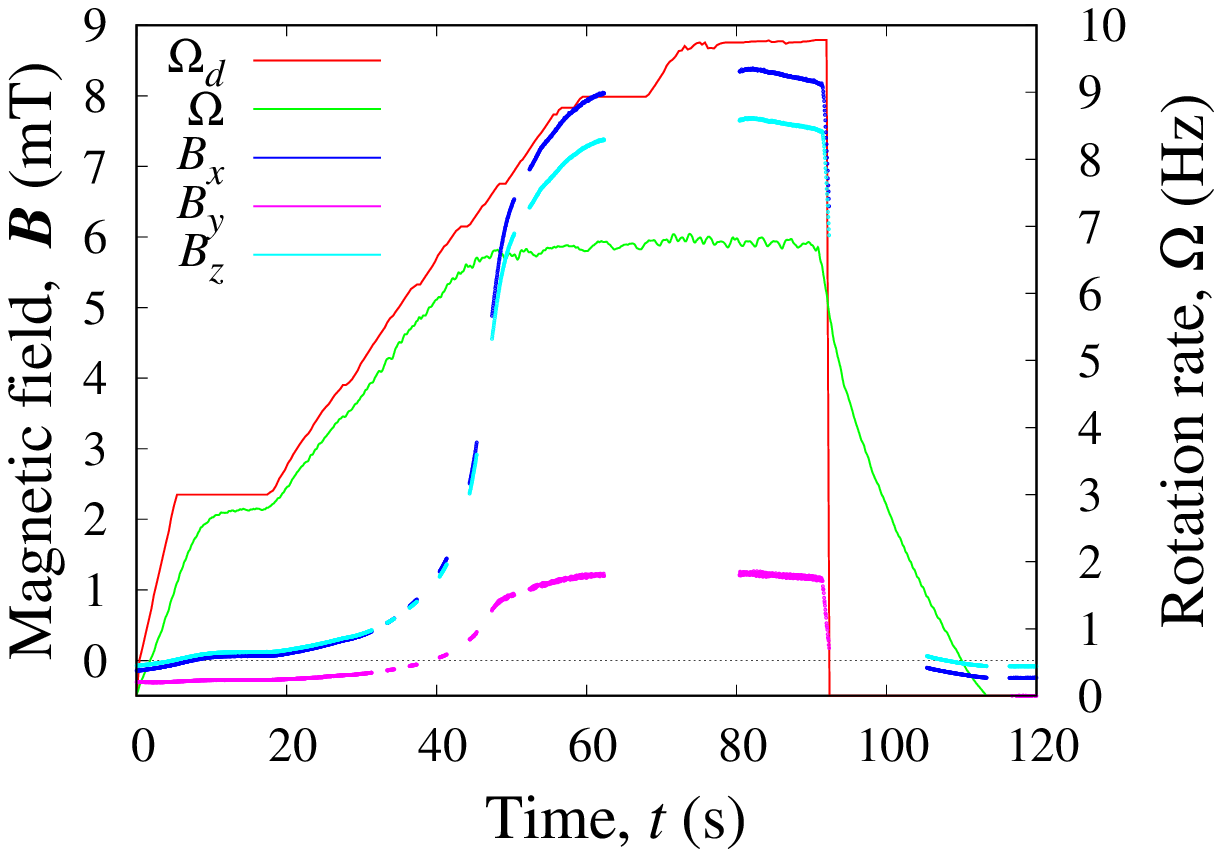}\put(-185,5){(b)}
\par\end{centering}
\caption{\label{fig:Bout}The magnetic field $(B_{x},B_{y},B_{z})$ at the
outer radius of the coil along with the driving $(\Omega_{d})$ and
rotation $(\Omega)$ frequencies in the second (a) and third (b) runs.}
\end{figure}

As the rotation rate increased, the magnetic field raised steeply,
especially its vertical $(y)$ component, which is seen Fig. \ref{fig:tim-ser}(a)
to exceed the upper limit of the low-field Hall sensor by reaching
$B_{y}\approx-\unit[9.3]{mT}$ at $\Omega\approx\unit[6.5]{Hz}$ $(t\approx\unit[76]{s})$.
A few seconds later, the increase in the driving frequency was halted
at $\Omega_{d}=\unit[7.33]{Hz.}$ At this point, the rotation rate,
which had reached $\Omega\approx\unit[6.7]{Hz},$ started to fall
while the $x$ and $z$ components of the magnetic field kept growing
for a few more seconds. The same held also for the coil voltage which
can be seen in Fig. \ref{fig:tim-ser}(b) to rise from $V\approx\unit[7.7]{mV}$
at the velocity maximum to $V_{\text{max }}\approx\unit[10]{mV}$
at $t\approx\unit[82]{s}.$ However, this time delay between the rotation
rate and the other two measured quantities was too short to be determined
reliably. Namely, it was comparable to the uncertainty in the time
synchronization between the measurements of these quantities which
were recorded using different devices.

The slowdown means that the electromagnetic braking torque acting
on the disc had shot past the equilibrium point corresponding to the
driving torque produced by the motor. As discussed before, this effect
can be due to transient eddy currents which, according to the previous
estimates, decay over the characteristic time $\sim$$\unit[1]{s}.$
They reduce the magnetic flux through the disc and so the associated
electromagnetic braking torque when the disc accelerates while the
opposite is the case when the disc decelerates. Although the decrease
in the rotation rate at the fixed driving frequency and voltage results
in the rise of the current running through the motor (seen in Fig.
\ref{fig:tim-ser}c), it does not prevent the slowdown because the
maximal torque the motor can develop is lower than the electromagnetic
braking torque. Actually, the deceleration is escalated further by
the relatively large difference between the driving and rotation frequencies.
When the frequency difference becomes too large, the motor breaks
down because the output torque, which is described by Eq. (\ref{eq:torque}),
starts dropping with the rotation rate. 

Although, as seen in Fig. (\ref{fig:tim-ser}c), the coil voltage
drops together with the rotation frequency, the rate of decrease is
lower than that at which the voltage increased when the disc accelerated.
Both the voltage and the magnetic field are noticeably higher than
the respective values at the same rotation rate in the acceleration
stage. This difference is likely due to the remanent magnetization
of the iron frame. However, the transient eddy currents, which delay
the variation of the magnetic flux through the disc as discussed above,
can have a similar effect.  

\begin{figure}[h]
\begin{centering}
\includegraphics[width=0.5\columnwidth]{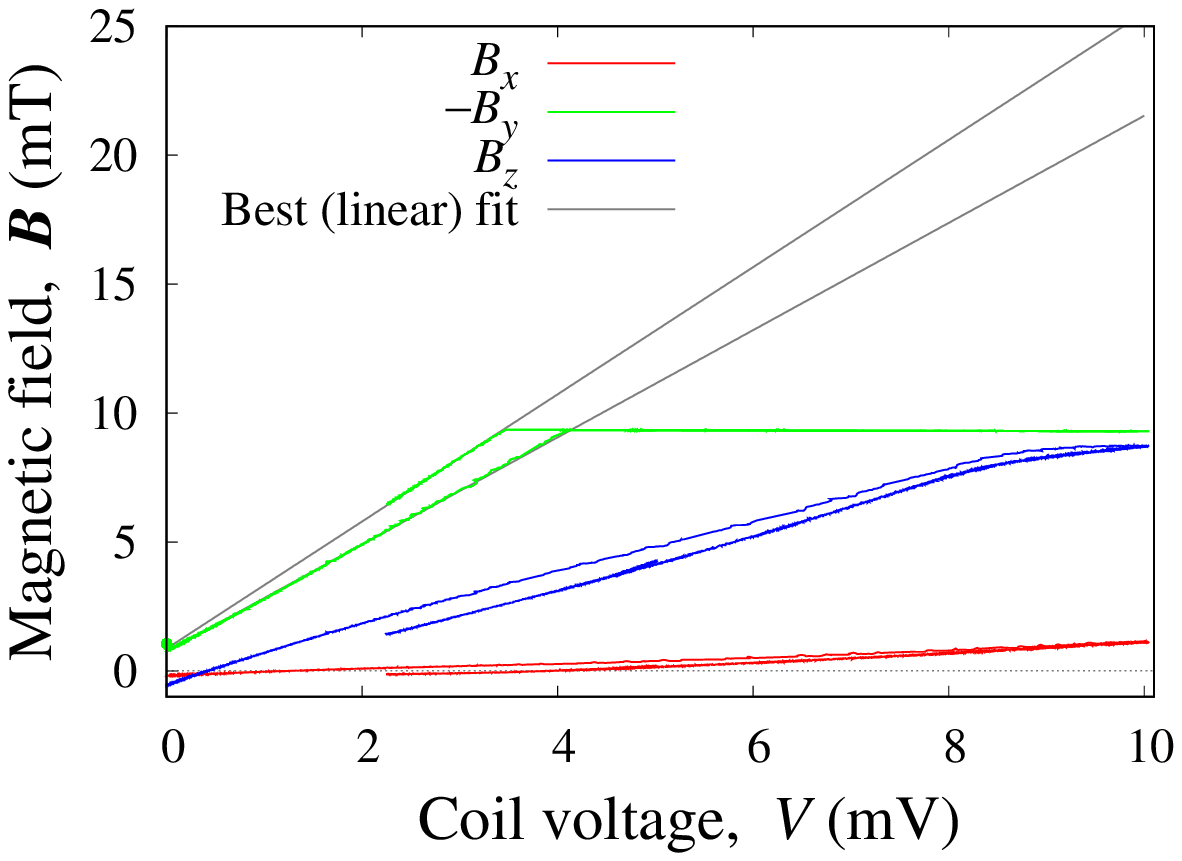}\put(-185,5){(a)}\includegraphics[width=0.5\columnwidth]{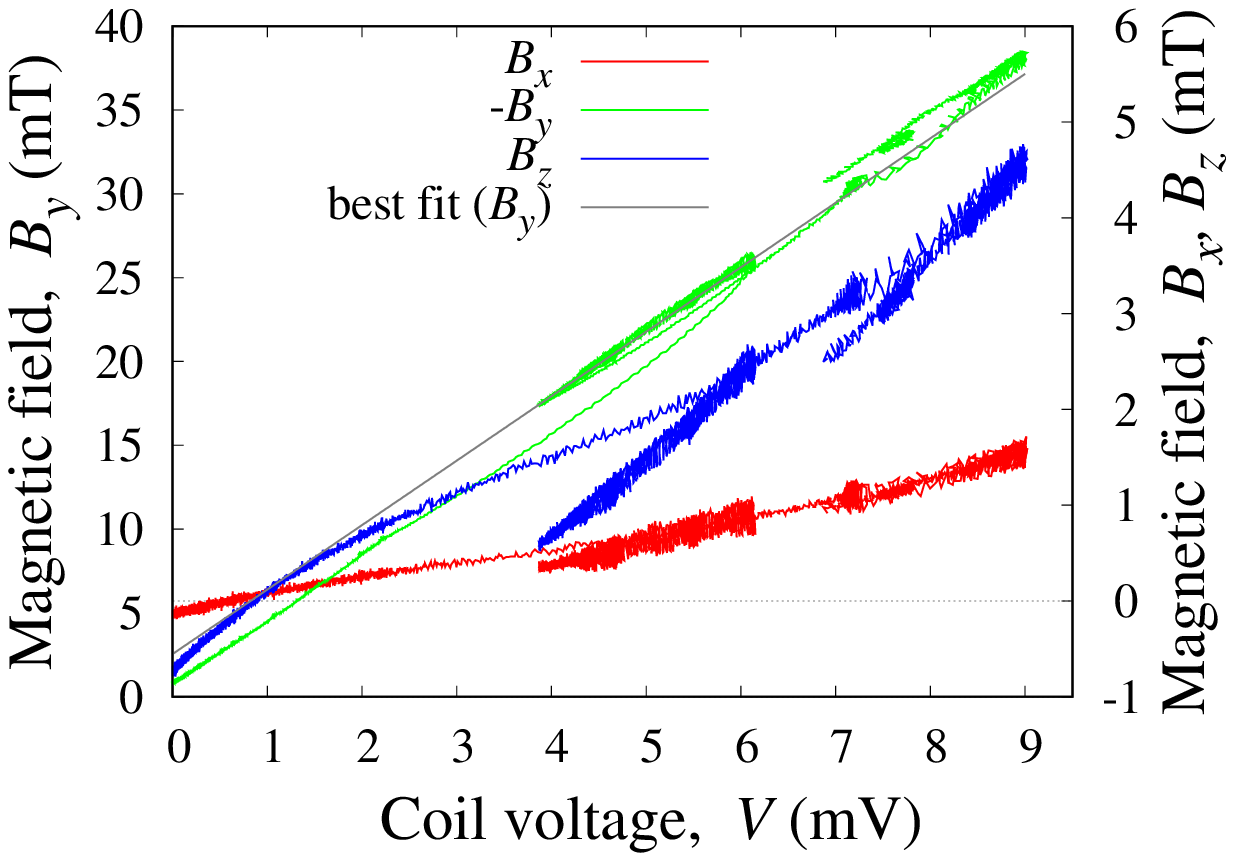}\put(-185,5){(b)}
\par\end{centering}
\begin{centering}
\includegraphics[width=0.5\columnwidth]{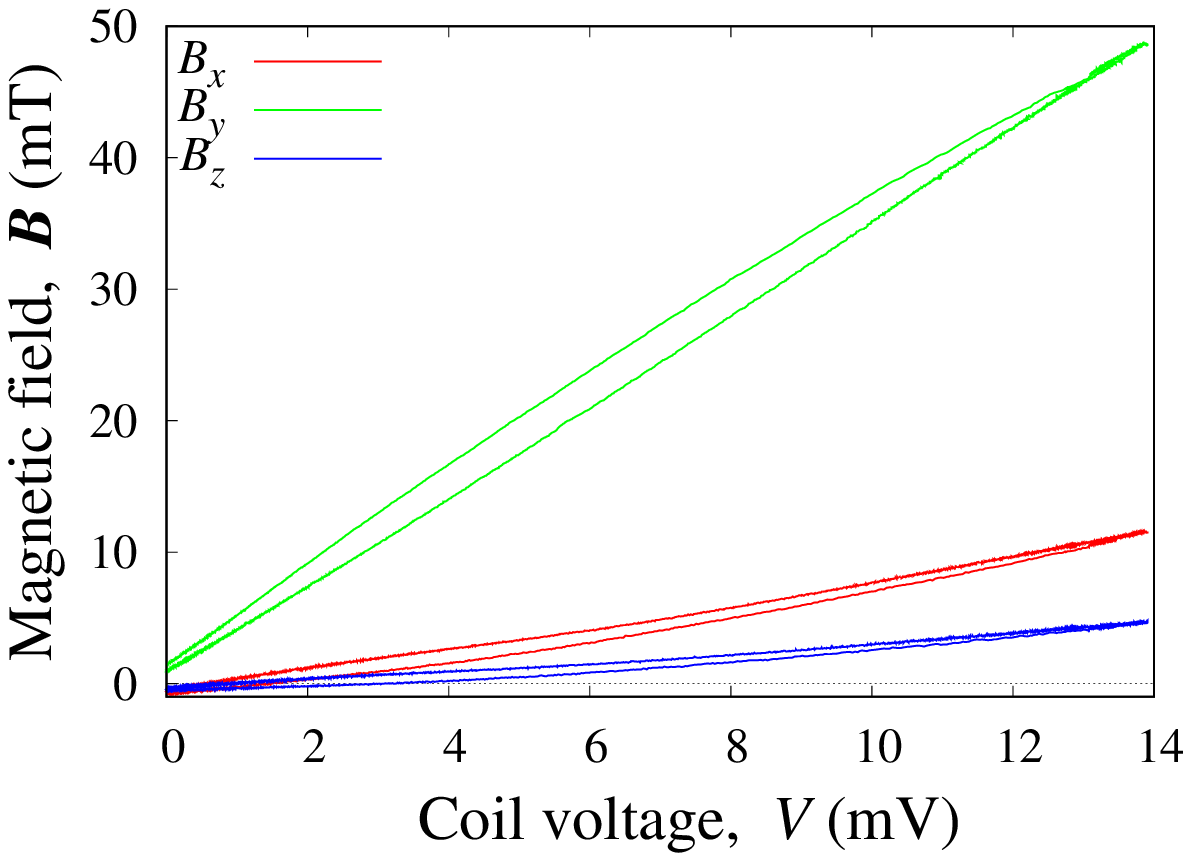}\put(-185,5){(c)}\includegraphics[width=0.5\columnwidth]{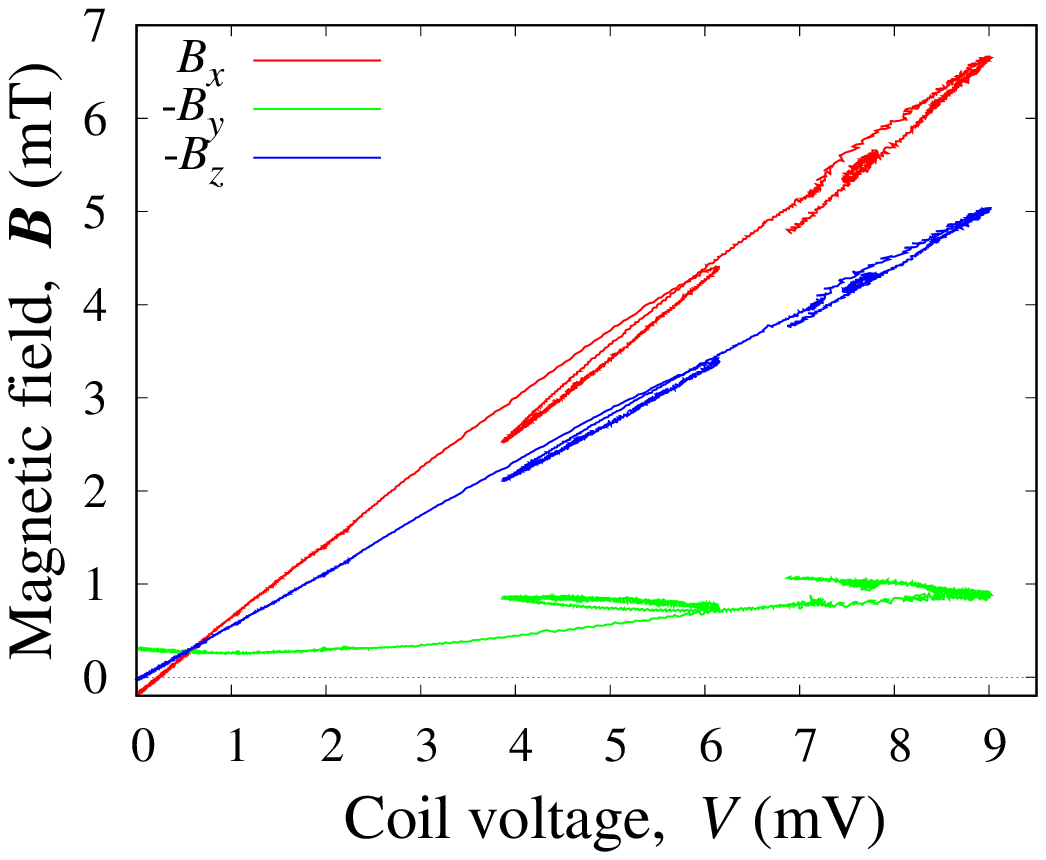}\put(-185,5){(d)}
\par\end{centering}
\caption{\label{fig:Emf-B}The magnetic field at the inner radius of spiral
slits for three runs (a,b,c) and at the outer radius for the second
run (d) versus the coil voltage.}
\end{figure}

Attempting to reduce the slip, at $t\approx\unit[126]{s},$ we started
to ramp down $\Omega_{d}$ to $\unit[7]{Hz}$ which was reached in
$\approx\unit[7]{s}.$ This, however, just increased the slowdown
rate without noticeably reducing the slip. The subsequent reduction
of $\Omega_{d}$ to $\unit[6.7]{Hz}$ had a similar effect. The slowdown
temporarily reversed at $t\approx\unit[156]{s}$ when $\Omega_{d}$
was ramped up to the original level of $\unit[7.33]{Hz.}$ Acceleration
lasted only for $\unit[\approx15]{s}$ after which $\Omega$ resumed
falling. When, at $t\approx\unit[192]{s},$ $\Omega_{d}$ was ramped
down in $\approx\unit[6]{s}$ to $\unit[6.73]{Hz},$ the falling rate
of rotation frequency as well as that of the magnetic field initially
increased. After $\approx\unit[6]{s},$ when the rotation rate and
the $y$ component of the magnetic field had dropped down to $\Omega\approx\unit[1.6]{Hz}$
and $B_{y}\approx\unit[5.6]{mT},$ the slowdown stopped and the disc
started to re-accelerate. Unfortunately, a few seconds later, the
liquid metal contacts started to fail due to the excessive oxidation
caused by the exposure to the air. The failure started with sporadic
current interruptions which appear as irregular oscillations of the
magnetic field and coil voltage in Figs. \ref{fig:tim-ser}(a,b).
The electric contact vanished completely in $\approx\unit[8]{s}$
and so did the associated electromagnetic braking torque. As a result,
the disc quickly accelerated to the prescribed VFD frequency, and
the motor current dropped from $\approx\unit[21]{A}$ to $\unit[5.7]{A}$
(see Fig. \ref{fig:tim-ser}c). The motor was turned off at $t\approx\unit[244]{s}$
and slowed down by the VFD linearly to a complete stop in $\approx\unit[60]{s}.$
The magnetic field measured at this point, $\vec{B}_{0}\approx\unit[-(0.23,1.0,0.72)]{mT,}$
was $26-36\%$ higher than at the beginning of the run. This difference
is obviously due to the remanent magnetization of the iron frame.

The peak value of $B_{y},$ which was not measured because it exceeded
the upper detection limit of the low-field Hall probe, can be estimated
by extrapolating its variation with the coil voltage. In a quasi-stationary
regime, both the voltage and the associated magnetic field are expected
to vary linearly with the current. This is confirmed by their mutual
dependence plotted in Fig. \ref{fig:Emf-B}(a). Extrapolating the
rising part of $V$and $B_{y},$ which corresponds to the lower branch,
we have $B_{y}\approx\unit[22]{mT}$ at the maximal coil voltage $V_{\text{ }}\approx\unit[10]{mV.}$
Extrapolating the falling part, which corresponds to the upper branch,
we respectively have $B_{y}\approx\unit[25]{mT.}$ The dependence
of the magnetic field on its direction of variation is most likely
due to the remanent magnetization of the iron frame. 

It is interesting to note that the relation between the coil voltage
and the magnetic field is nearly linear except the $z$ component
which appears to saturate close to $\unit[9]{mT,}$i.e., the upper
detection limit of the low-field Hall probe. Also note that both the
iron frame and the spiral slits made the magnetic field strongly non-uniform
at the inner and outer edges of the coil where the probes were placed.
As a result, the strength, as well as the direction of the magnetic
field, were found to vary noticeably with the location of the Hall
sensor which was not fixed and may vary between the runs. 

In the second run, the low-field probe was moved to the outer radius
and the medium-field probe was placed near the inner radius of the
coil. The driving frequency was ramped up through intermediate steps
rather than continuously as in the first run. As seen in Fig. \ref{fig:tim-ser}(d,e),
this resulted in the rise of the magnetic field and the coil voltage
at an increasing rate. A particularly steep increase was observed
when $\Omega_{d}$ was ramped up from $\unit[7]{Hz}$ at $\unit[t\approx194]{s}$
to $\unit[7.67]{Hz}$ at $\unit[t\approx200]{s}.$ As the rotation
rate increased from $\unit[6.1]{Hz}$ to $\unit[6.4]{Hz},$ $B_{y}$at
the inner radius rose steeply from $\unit[9.5]{mT}$ up to $\unit[26.5]{mT.}$
The magnetic field at the outer radius can be seen in Fig. \ref{fig:Bout}
to increase in a similar way. The coil voltage, which is plotted Fig.
\ref{fig:tim-ser}(e), increased respectively from $\unit[2.2]{mV}$
to $\unit[6.1]{mV.}$ After that the rotation rate started to fall
and so did the magnetic field along with the coil voltage. It indicates
that the electromagnetic braking torque had again exceeded the torque
that the motor can produce at the given parameters. Thus, the motor
was not able to sustain the previously attained field strength and
went into the breakdown regime. The rotation rate kept falling until
$\Omega\approx\unit[5.6]{Hz}$ was reached at $\unit[t\approx255]{s}.$
At this point, the slowdown ended and the rotation rate started to
increase slowly. This increase lasted only for $t\approx\unit[25]{s}$
after which the rotation rate dropped again reaching $\Omega\approx\unit[5.5]{Hz}$
at $\unit[t\approx297]{s}.$ The subsequent increase in the rotation
rate was enhanced at $\unit[t\approx304]{s}$ by ramping up the driving
frequency to $\unit[8.47]{Hz}$ which was reached in $\unit[\approx6]{s}.$
As a result, the rotation rate raised forming a plateau at $\unit[t\approx315]{s}$
with $\Omega\approx\unit[6.3]{Hz}$ which lasted for $\unit[\approx8]{s}.$
The vertical magnetic field, which is shown in Fig. \ref{fig:tim-ser}(d),
reached $B_{y}\unit[\approx30]{mT}$ at $\unit[t\approx319]{s}$ and
stayed close to this value for $\unit[\approx6]{s}.$ After the plateau,
the rotation rate can be seen to increase slightly reaching $\approx\unit[6.5]{Hz}$
in $\unit[\approx2]{s}.$ This caused a steep rise in the magnetic
field which reached a peak of $\unit[\approx38]{mT}$ at $\unit[t\approx333]{s}.$
As before, the motor could no longer sustain the attained rotation
rate which, thus, started to fall closely followed by the magnetic
field until reaching $\Omega\approx\unit[6.1]{Hz}$ and $B_{y}\unit[\approx32.5]{mT}$
at $\unit[t\approx359]{s}.$ At this point, we started to ramp down
the driving frequency thus disrupting the subsequent growth phase.
At $\unit[t\approx386]{s},$ when the rotation rate had slowed down
to $\Omega\approx\unit[5.2]{Hz},$ oscillations in the magnetic field
increased abruptly and the field started to fall sharply. The same
may be seen in Fig. \ref{fig:tim-ser}(e) to happen to the coil voltage.
This introduces the breakdown of electric contact which was completely
lost at $\unit[t\approx400]{s}.$ At this point, the magnetic field
dropped down close to its initial background value. As the electromagnetic
braking torque vanished, the disc rotation rate increased approaching
the VFD driving frequency. Although the motor closely followed the
driving frequency in this regime, the VFD caused some oscillations
in the rotation rate around $\Omega\approx\unit[7.5]{Hz}$ which may
be seen in Figs. \ref{fig:tim-ser}(d-f).

In the third run, which is documented in Figs. \ref{fig:tim-ser}(g-j),
the driving frequency $\Omega_{d}$ was set higher and ramped up faster
than in the second run. Firstly, the driving frequency was ramped
up in $\unit[5.4]{s}$ to $\Omega_{d}\approx\unit[3]{Hz}.$ Secondly,
after $\unit[\approx12]{s},$ when the rotation rate had stabilized
at $\Omega\approx\unit[2.8]{Hz},$ the driving frequency was ramped
up to $\Omega_{d}\approx\unit[8.9]{Hz,}$ which was reached nearly
linearly via a few short intermediate steps at $\unit[t\approx60]{s}.$
The motor rotation rate also increased nearly linearly but at a slightly
lower pace than the driving frequency and only up to $\Omega\approx\unit[6.7]{Hz}$
which was attained at $\unit[t\approx45]{s}.$ At this point, $\Omega$
ceased to increase further with $\Omega_{d}.$ The subsequent results
show that only the magnetic field was affected by $\Omega_{d}$ while
$\Omega$ stayed nearly constant. Namely, the magnetic field, which
saturated at $B_{y}\unit[\approx45]{mT}$ when the driving frequency
was set to $\Omega_{d}\approx\unit[8.9]{Hz,}$can be seen in Fig.
\ref{fig:tim-ser}(g) to reach $B_{y}\unit[\approx48.6]{mT}$ at $\unit[t\approx74]{s}$
when the driving frequency was ramped up $\Omega_{d}\approx\unit[9.7]{Hz}.$
Although $B_{y}$ fell off subsequently, it remained well above the
previous equilibrium level until the motor was switched off at $\unit[t\approx92]{s}.$

\section{\label{sec:model}An extended disc dynamo model}

In this section, we extend our original quasi-stationary disc dynamo
model \citep{Priede2013} by including several effects which are essential
for the interpretation and analysis of the experimental results. In
order to capture the temporal evolution and saturation of the disc
dynamo, first of all, we need to take into account two transient inductive
effects. The first is an additional potential difference between the
inner and outer rim of the coil which is induced by a time-dependent
electric current $I_{1}(t)$ flowing along the spiral arms of the
coil and then connecting radially through the disc. Owing to the fixed
distribution of $I_{1}(t)$ in the coil, this voltage can be expressed
analytically in terms of the rate of variation of the associated magnetic
flux (Eq. (8) in \citep{Priede2013}). The second effect, which is
neglected in the original disc dynamo model \citep{Bullard1955},
is the eddy current induced in the disc by a changing magnetic field
\citep{Moffatt1979}. The linear density of this purely azimuthal
current is governed by Ohm's law
\[
J_{2}(r,t)=\bar{\sigma}E_{\phi},
\]
where $\bar{\sigma}=\sigma d$ is the effective electric conductivity
of the disc and $E_{\phi}=-\partial_{t}A_{\phi}$ is the induced electric
field which follows from Faraday's law with $A_{\phi}$ standing for
the azimuthal component of the vector potential. Using the Biot-Savart
law, the latter can be written as follows
\begin{equation}
A_{\phi}(r,t)=\frac{\mu_{0}}{4\pi}\intop_{\tilde{r}_{i}}^{r_{o}}\intop_{0}^{2\pi}\frac{(J_{1}(r',t)-\bar{\sigma}\partial_{t}A_{\phi})r'\cos\phi}{\sqrt{r^{2}+r'^{2}-2rr'\cos\phi}}\thinspace\mathrm{d}\phi\thinspace\mathrm{d}r',\label{eq:Aphi}
\end{equation}
which is an integro-differential equation governing the evolution
of $A_{\phi}$ over the radius of the system; $J_{1}(r,t)=\beta I_{1}(t)/2\pi r$
stands for the azimuthal current distribution in the coil and $\mu_{0}=\unit[4\pi\times10^{-7}]{H/m}$
is the permeability of vacuum. Note that the vertical gap between
the disc and the coil is supposed to be much smaller than the radial
size. Thus, the magnetic flux passing through the disc is approximately
the same as that passing through the coil \citep{Priede2013}. In
the following, instead of solving Eq. (\ref{eq:Aphi}), which is numerically
complicated, we pursue a simplified approach by assuming that the
aforementioned magnetic flux can be written as 
\begin{equation}
\Phi=L_{1}I_{1}+L_{2}I_{2}+\Phi_{0},\label{eq:flux}
\end{equation}
where the first two terms on the RHS represent the fluxes generated
by the coil and the disc, respectively, and $\Phi_{0}$ is a background
flux due to external magnetic which may be Earth's magnetic field
as well as to that produced by the magnetized iron frame. This model
differs from that of Moffatt \citep{Moffatt1979} first, by taking
into account the external field and, second, by neglecting the leakage
of the magnetic flux through the gap between the coil and the disc.

Applying Kirchhoff's voltage law to the primary dynamo circuit, which
carries the current $I_{1},$ and to the disc, which constitutes the
secondary circuit, we obtain 
\begin{eqnarray}
\dot{\Phi}+R_{1}I_{1} & = & \Omega\Phi,\label{eq:I1}\\
\dot{\Phi}+R_{2}I_{2} & = & 0,\label{eq:I2}
\end{eqnarray}
where $R_{1}$ and $R_{2}$ are the effective ohmic resistances of
respective circuits. The system is completed by the angular momentum
equation:
\begin{equation}
\mathcal{I}\dot{\Omega}=G-I_{1}\Phi-k\Omega,\label{eq:dyn}
\end{equation}
where $\mathcal{I}$ is the moment of inertia of the disc, $G$ is
the driving torque produced by the motor and $I_{1}\Phi$ is the electromagnetic
braking torque \citep{Moffatt1979}. The last term with the coefficient
$k$ represents a hyrodynamic-type friction. It can also model the
variation of the driving torque with the rotation speed as in asynchronous
electric motor controlled by the VFD with a fixed voltage and frequency
ratio. The system can be reduced to two equations for $\Phi$ and
$\Omega$ as follows. First, substituting $I_{2}=-\dot{\Phi}/R_{2}$
from Eq. (\ref{eq:I2}) into Eq. (\ref{eq:flux}) we have 
\[
I_{1}=(\Phi-\Phi_{0}+\dot{\Phi}L_{2}/R_{2})/L_{1},
\]
Second, substituting this expression for $I_{1}$ into Eqs. (\ref{eq:I1})
and (\ref{eq:dyn}), after a few rearrangements, we obtain:
\begin{eqnarray}
\tau_{0}\dot{\Phi} & = & \Phi(\tau_{1}\Omega-1)+\Phi_{0}\label{eq:Phi}\\
\mathcal{I}\dot{\Omega} & = & G-\Phi(\Phi+\tau_{2}\dot{\Phi})/L_{1}-k\Omega,\label{eq:Omg}
\end{eqnarray}
where $\tau_{i}={\textstyle L_{i}/R_{i}}$ are the characteristic
electromagnetic decay times (the time constants) for the primary $(i=1)$
and secondary $(i=2)$ circuits and $\tau_{0}=\tau_{1}+\tau_{2}$.
Note that the kinematic dynamo problem, which corresponds to a prescribed
rotation rate $\Omega,$ is posed just by Eq. (\ref{eq:Phi}). According
to this equation, a small initial current perturbation grows exponentially
with time if $\Omega>\Omega_{c},$ where 
\begin{equation}
\Omega_{c}=\tau_{1}^{-1}\label{eq:omgc}
\end{equation}
 is a critical rotation rate defining the dynamo threshold. Scaling
rotation rate and time with $\Omega_{c}$ and $\tau_{0},$ respectively,
Eqs. (\ref{eq:Phi}) and (\ref{eq:Omg}) can be written in the dimensionless
form as follows 
\begin{eqnarray}
\dot{\Phi} & = & \Phi(\Omega-1)+\Phi_{0},\label{eq:Phi-ND}\\
\dot{\Omega} & = & \Gamma(1-\kappa\Omega)-\Phi\left(\Phi-\Phi_{0}-\lambda\dot{\Phi}\right),\label{eq:Omg-ND}
\end{eqnarray}
were $\Gamma$ is a dimensionless driving torque, $\kappa$ is a friction
coefficient defined relative to $\Gamma$ and $\lambda=\tau_{2}/(\tau_{1}+\tau_{2})$
is a parameter defining the relative significance of eddy currents
in the disc. In the following, we assume mechanical friction to be
negligible, i.e., $\kappa\ll\Gamma,$ and use $\kappa=0$ unless
stated otherwise. Finally, using Eq. (\ref{eq:Phi-ND}) to eliminate
$\dot{\Phi}$ on the RHS of Eq. (\ref{eq:Omg-ND}), we obtain a second-order
dynamical system. 

If the disc is segmented by spiral slits like the coil, or by radial
insulating strips as suggested by Moffatt \citep{Moffatt1979}, so
that no azimuthal eddy currents can circulate in the disc, we have
$\lambda=0.$ In this case, our model, like that of Moffatt \citep{Moffatt1979},
reduces to the original Bullard dynamo model \citep{Bullard1955}.
If the disc is solid and well conducting, as in our experiment, so
that $\tau_{2}\gg\tau_{1},$ which will be shown in the following,
we have $\lambda=1.$ 

The driving torque $\Gamma$ can conveniently be defined using the
dynamic response time $\tau_{d},$ i.e., the time required by the
motor to spin up the disc to the critical rotation rate (\ref{eq:omgc}).
In the case of constant driving torque and dominating inertia, Eq.
(\ref{eq:Omg-ND}) yields $\Omega=\Gamma t.$ Taking into account
that $\Omega_{c}=1,$ we have $\Gamma=(\tau_{d}/\tau_{0})^{-1},$
which is a dimensionless counterpart of $\tau_{d}^{-1}$. Similarly,
in the case of a constant power $\Pi=\Omega\Gamma$, Eq. (\ref{eq:Omg-ND})
yields $\frac{1}{2}\Omega^{2}=\Pi t$ and, thus, we have $\Gamma=\frac{1}{2}(\Omega\tau_{d}/\tau_{0})^{-1}.$

For a disc of radius $r_{o}$ and thickness $d\ll r_{o},$ which can
be treated as a thin sheet with the effective electric conductivity
$\bar{\sigma}=\sigma d,$ the characteristic time over which eddy
currents decay can be estimated as $\tau_{2}\sim\mu_{0}\sigma dr_{o}.$
For our set-up with $r_{o}=10d=\unit[0.3]{m}$ and $\sigma=\unit[58.5]{S/m},$
the electromagnetic time constant of the secondary circuit is $\tau_{2}\approx\unit[0.7]{s}.$ 

The time constant of the primary circuit $\tau_{1}$ can be estimated
using Eq. (\ref{eq:omgc}) and the corresponding critical magnetic
Reynolds number, which can be written as follows
\global\long\def\Rm{\mathit{Rm}}%
\[
\Rm_{c}=\mu_{0}\sigma dr_{o}\Omega_{c}=\tau_{2}/\tau_{1}.
\]
Then taking into account that $\Rm_{c}\approx40$ \citep{Priede2013},
we have $\tau_{2}=\tau_{1}\Rm_{c}\gg\tau_{1}$ and, thus, $\lambda\approx1.$

\begin{figure}
\begin{centering}
\includegraphics[width=0.5\columnwidth]{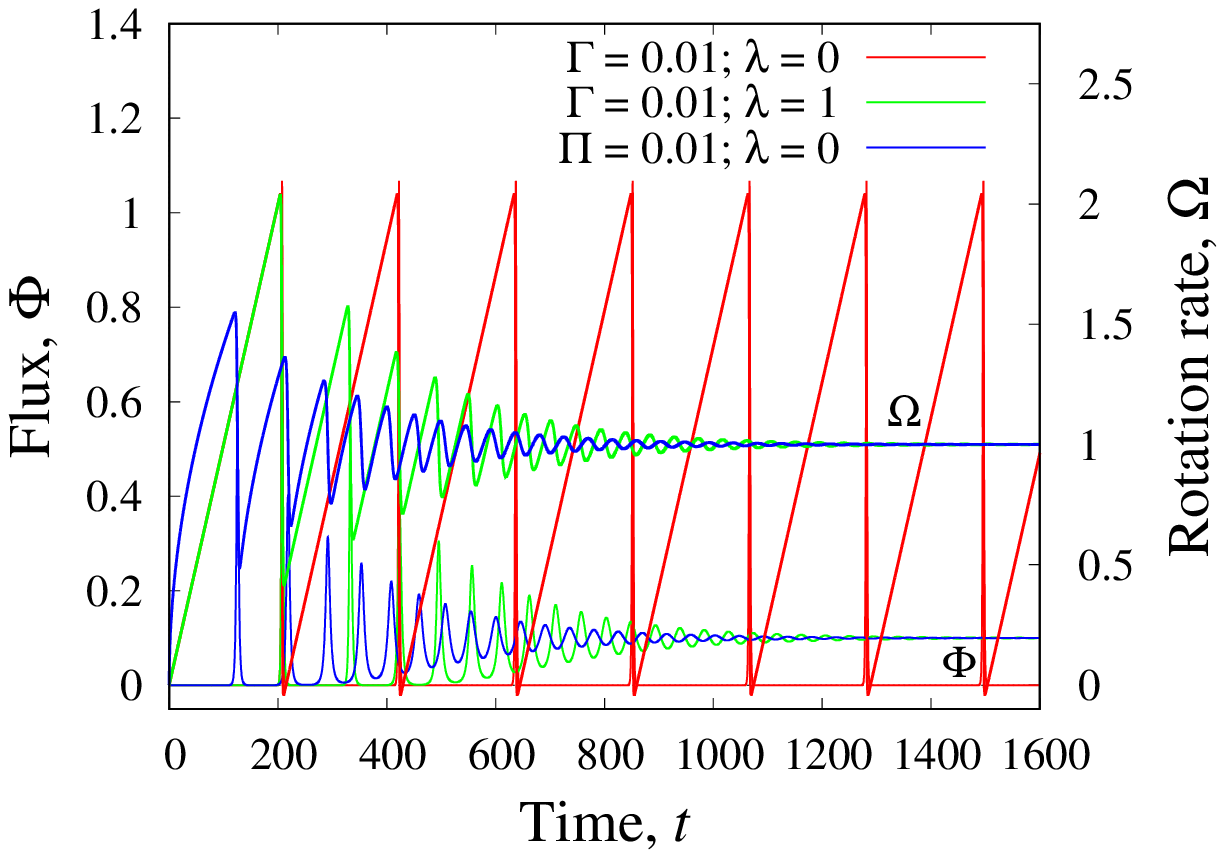}\put(-185,5){(a)}\includegraphics[width=0.5\columnwidth]{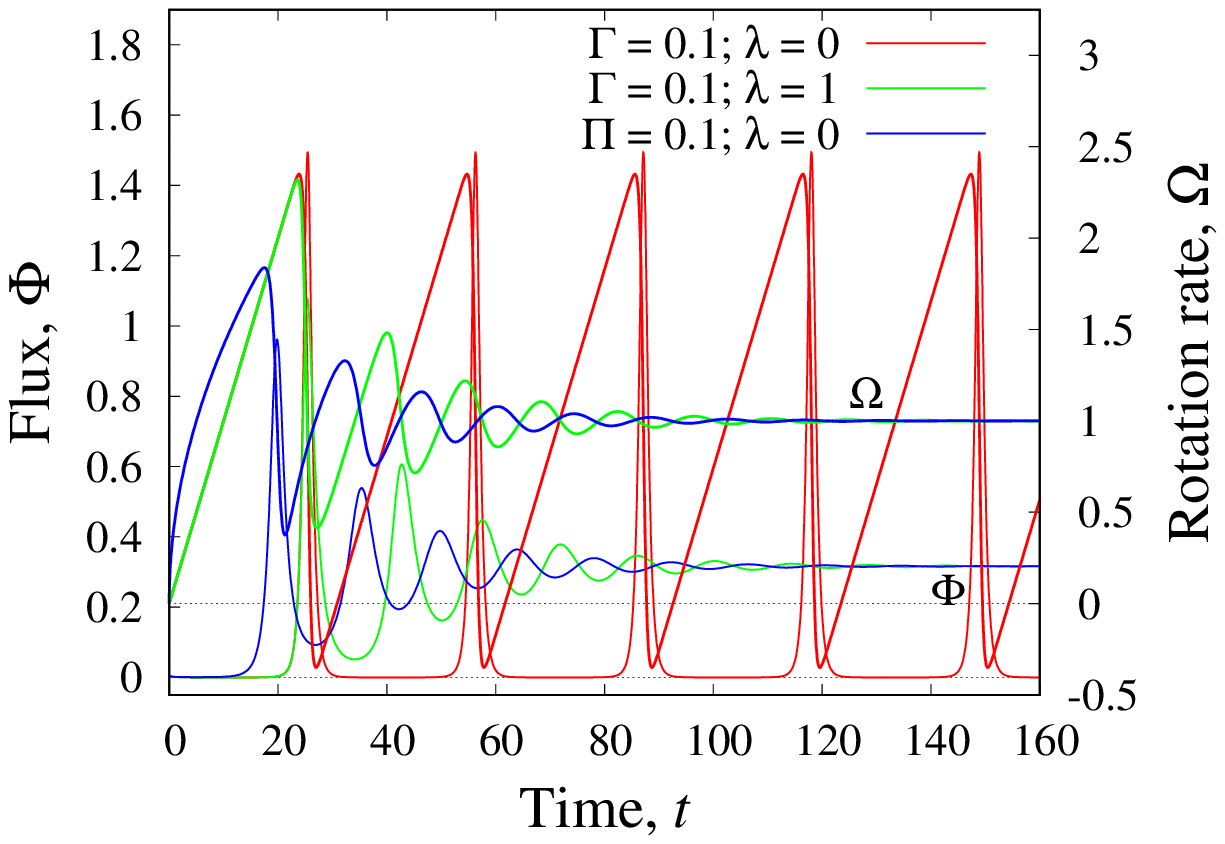}\put(-185,5){(b)}
\par\end{centering}
\caption{\label{fig:tm_phi0}Temporal evolution of the rotation rate $\Omega$
and the magnetic flux $\Phi$ with a small initial value and no background
magnetic field $(\Phi_{0}=0)$ and no mechanical friction $(\kappa=0)$
when the disc is driven either by a constant torque $\Gamma=0.1\,\text{(a)},0.01\,\text{(b)}$
or a constant power $\Pi=0.1\,\text{(a)},0.01\,\text{(b);}$ $\lambda$
characterizes the effect of eddy currents in the disc which is negligible
when $\lambda=0$ and significant when $\lambda=1.$}
\end{figure}

The characteristic mechanic response time in which the disc attains
the expected critical rotation frequency $f_{c}\approx\unit[10]{Hz}$
when the motor is run at $P\approx\unit[1]{kW,}$ i.e. half of its
rated power, can be estimated as 
\[
\tau_{d}=\frac{1}{2}\mathcal{I}\Omega_{c}^{2}/P\approx\pi^{3}\rho dr_{o}^{4}f_{c}^{2}/P\approx\unit[7]{s},
\]
where $\rho=\unit[8.96\times10^{3}]{kg/m^{3}}$ is the density of
copper. This time being by an order of magnitude longer than the electromagnetic
response time $\tau_{1}$ corresponds to $\Gamma\sim0.1.$

There are two distinct ways how the dynamo can manifest itself. In
the classical scenario, the background field is assumed to be absent
$(\Phi_{0}=0).$ In this case, as soon as the rotation rate exceeds
$\Omega_{c}=1$, the magnetic field starts to grow exponentially in
time at the rate $\Omega-1$. This scenario, which corresponds to
an instability of the base state $\Phi=0$ at $\Omega>1,$ is captured
by Eqs. (\ref{eq:Phi-ND},\ref{eq:Omg-ND}) with the initial conditions
$\Omega(0)=0$ and $\Phi(0)=\Phi_{1},$ where $\Phi_{1}$ is a small
initial perturbation.

\begin{figure}
\begin{centering}
\includegraphics[width=0.5\columnwidth]{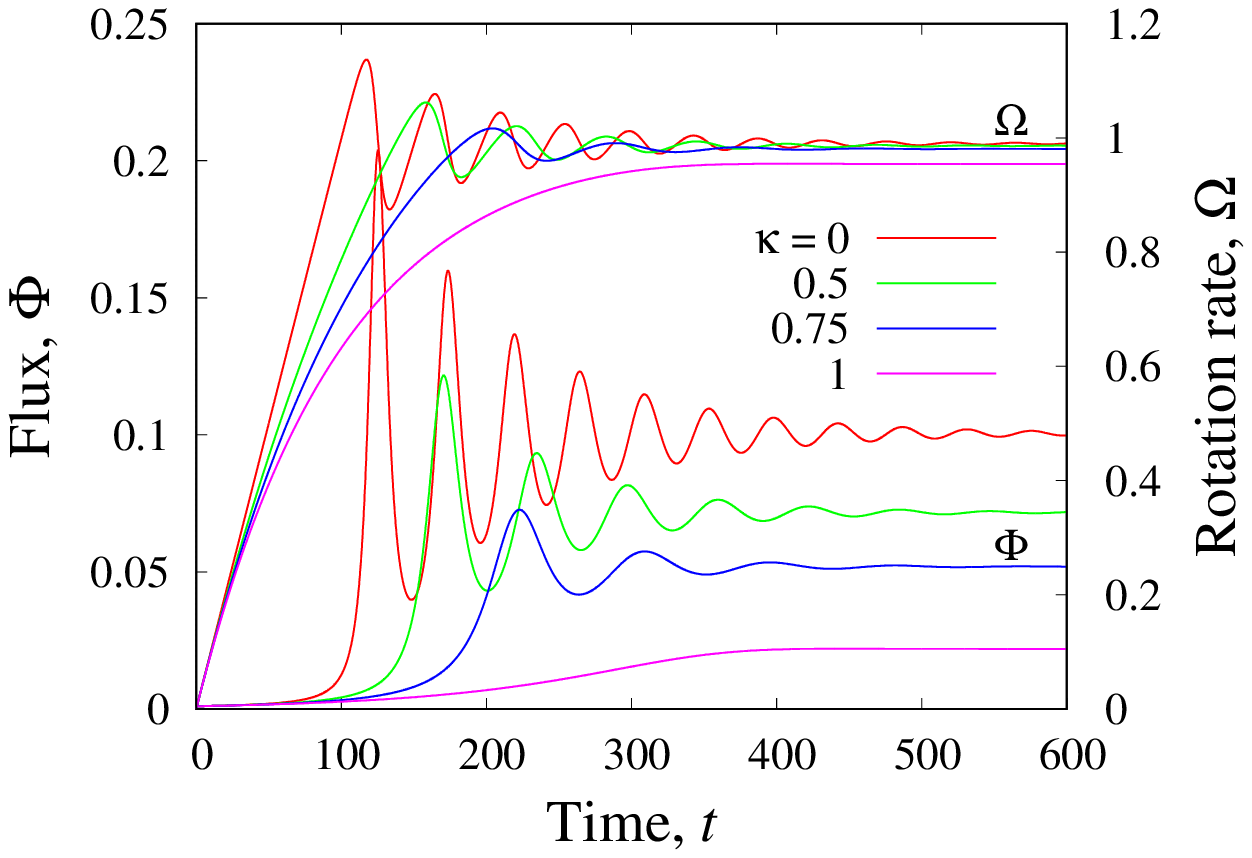}\put(-185,5){(a)}\includegraphics[width=0.5\columnwidth]{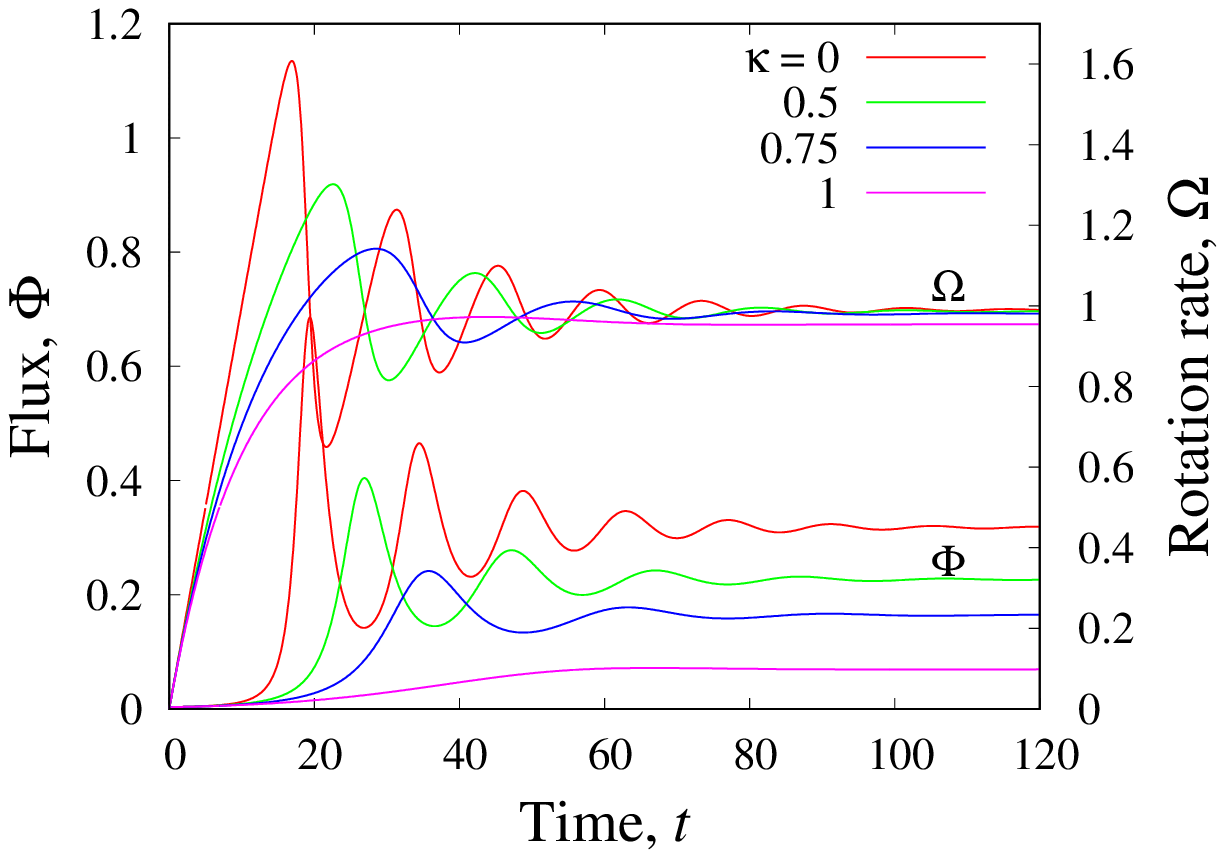}\put(-185,5){(b)}
\par\end{centering}
\caption{\label{fig:tm_phi.01}Temporal evolution of the rotation rate $\Omega$
and the magnetic flux $\Phi$ in the presence of a background field
with $\Phi_{0}=0.01\Gamma^{1/2},$ $\lambda=1,$ $\kappa=0,0.5,0.75,1$
when the disc is driven by a constant torque $\Gamma=0.01\,\text{(a)},0.1\,\text{(b).}$}
\end{figure}

The second scenario, which more adequately describes the experiment,
involves a non-zero background flux $\Phi_{0}$. In this case, Eq.
(\ref{eq:Phi-ND}) for a fixed rotation rate $\Omega$ has a stationary
solution $\Phi=\Phi_{0}/(1-\Omega),$ which shows that the magnetic
flux increases with the rotation rate as $\sim(1-\Omega)^{-1}$ and
becomes unbounded when $\Omega\rightarrow1.$ This is how the dynamo
manifests itself in the second scenario.

In reality, the growth of the magnetic field is limited by the electromagnetic
braking torque which at a certain point outweighs the driving torque.
The stationary solution for the flux defined by Eq. (\ref{eq:Omg-ND})
is 
\begin{equation}
\bar{\Phi}=\Phi_{0}/2+\sqrt{\Phi_{0}^{2}/4+\Gamma}.\label{eq:Phi-bar}
\end{equation}
The respective equilibrium rotation rate following from Eq. (\ref{eq:Phi-ND})
is $\bar{\Omega}=1-\Phi_{0}/\bar{\Phi}.$ If the background field
is negligible$,$ which is the case when $\Phi_{0}\ll\Gamma^{1/2},$
this equilibrium rotation rate becomes equal to the threshold value
$\Omega_{c}=1.$ It means that in this case, the driving torque changes
only the equilibrium magnetic flux $\bar{\Phi}=\Gamma^{1/2}$ but
not the disc rotation rate which stays equal to $\Omega_{c}=1$ \citep{Bullard1955}.
It is important to note, however, that in the Bullard model, which
corresponds to a constant driving torque $\Gamma$ with $\lambda=\Phi_{0}=0,$
this equilibrium is never reached and the dynamo keeps oscillating
around the equilibrium state unless it coincides with the initial
state \citep{Bullard1955}. Periodic oscillations of the Bullard dynamo
are due to an energy-like conserved quantity:
\[
(\Omega-1)^{2}+\Phi^{2}-\Gamma\ln\Phi^{2}=\text{const},
\]
 which makes Eqs. (\ref{eq:Phi-ND}, \ref{eq:Omg-ND}) integrable.
This is no longer the case when the dynamo is driven by a constant
power $\Pi=\Omega\Gamma,$ which also acts as a mechanical damping.
Namely, for small-amplitude oscillations around the equilibrium rotation
rate with $\Omega=1+\tilde{\Omega}$, we have 
\[
\Gamma=\Pi/\Omega\approx\Pi-\Pi\tilde{\Omega},
\]
where the term $\Pi$ represents the constant part of the driving
torque whereas the term $-\Pi\tilde{\Omega}$ is equivalent to the
damping with a coefficient equal to $\Pi$. There are two additional
electromagnetic damping effects in the modified Bullard dynamo model.
They become obvious when Eqs. (\ref{eq:Phi-ND}, \ref{eq:Omg-ND})
are combined into the following equation for small-amplitude magnetic
flux perturbation $\tilde{\Phi}=\Phi-\bar{\Phi}$ around the equilibrium
flux defined by Eq. (\ref{eq:Phi-bar}):

\[
\partial_{t}^{2}\tilde{\Phi}+\tilde{\kappa}\partial_{t}\tilde{\Phi}+(2\bar{\Phi}^{2}-\Phi_{0}\bar{\Phi})\tilde{\Phi}=0.
\]
This equation features an effective damping coefficient 
\begin{equation}
\tilde{\kappa}=\lambda\bar{\Phi}^{2}+\Phi_{0}/\bar{\Phi}\label{eq:ktld}
\end{equation}
with the two terms on the RHS being due to the eddy currents induced
in the disc and the background magnetic flux, respectively. 

The temporal evolution of the rotation rate and the magnetic flux
when the disc is driven either by a constant torque $\Gamma$ or power
$\Pi$ with no background magnetic field $(\Phi_{0}=0)$ and negligible
friction $(\kappa=0)$ is shown in Fig. \ref{fig:tm_phi0} for the
eddy current parameter $\lambda=0$ and $\lambda=1$. 

\begin{figure}
\begin{centering}
\includegraphics[width=0.5\columnwidth]{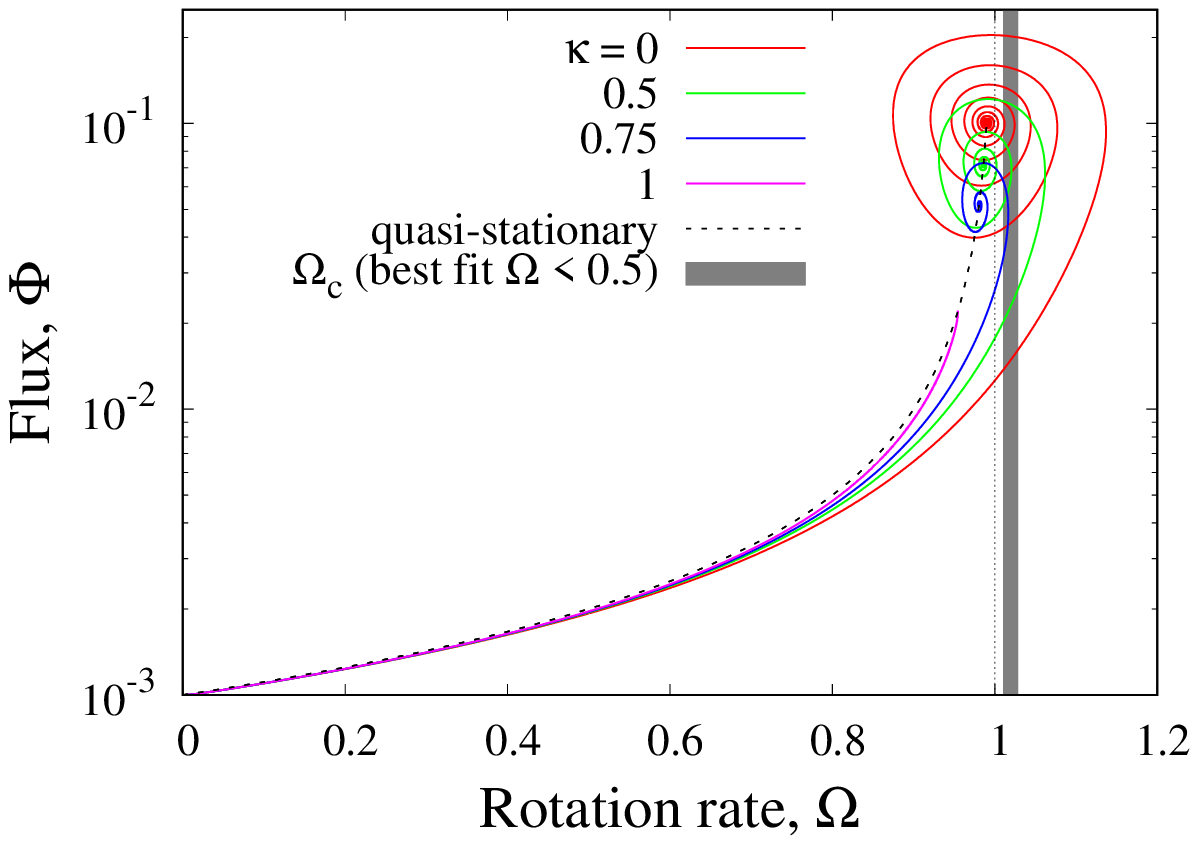}\put(-185,5){(a)}\includegraphics[width=0.5\columnwidth]{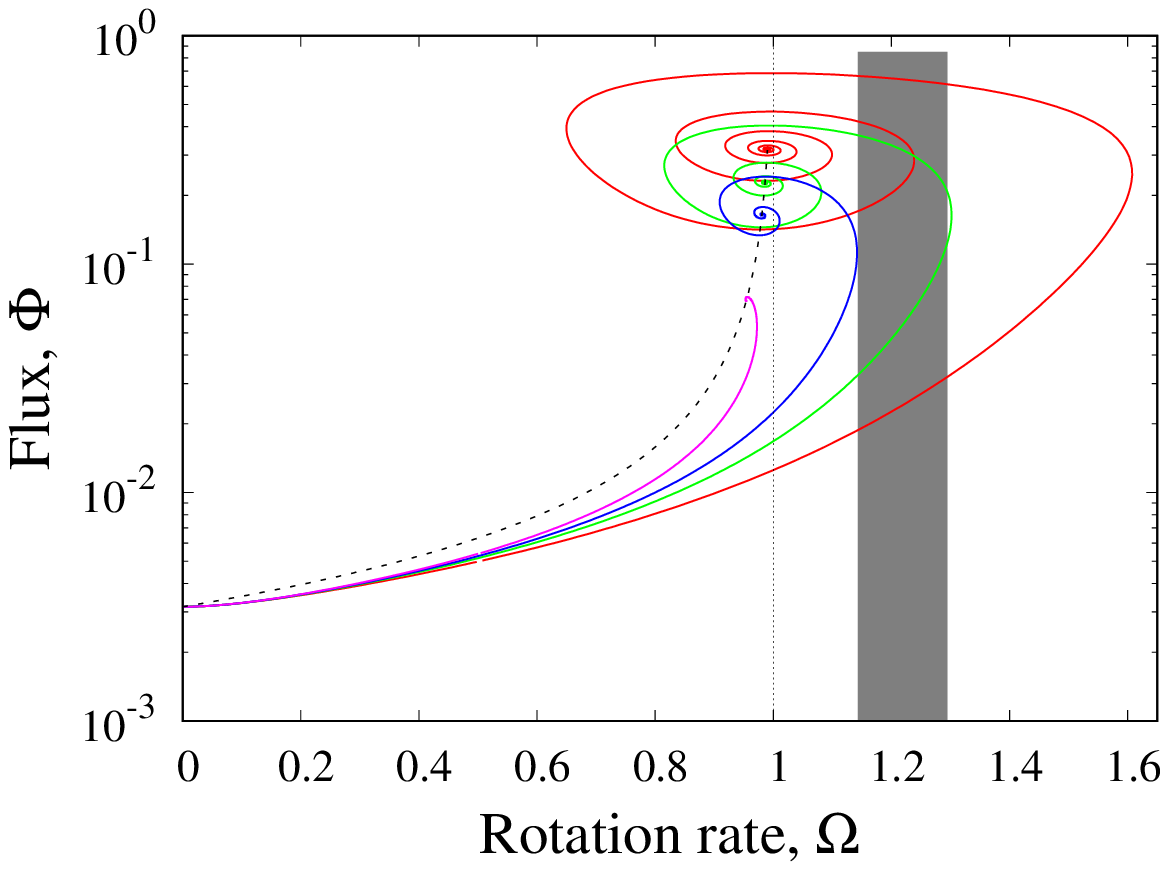}\put(-185,5){(b)}
\par\end{centering}
\caption{\label{fig:flx-omg}The magnetic flux $\Phi$ versus the rotation
rate $\Omega$ from Fig. \ref{fig:tm_phi.01} along with the quasi-stationary
solution (\ref{eq:Phi-ROT}) for $\Omega_{c}=1$ and $\Gamma=0.01\,\text{(a)},0.1\,\text{(b)}$.
The shaded strips show the range of $\Omega_{c}$ resulting from the
best fit of numerical results for various $\kappa$ and $\Omega\le\Omega_{c}/2$,
where the solution is relatively quasi-stationary.}
\end{figure}

The effect of a background magnetic field with $\Phi_{0}=0.01\Gamma^{1/2}$
on the evolution of the dynamo driven by a constant torque with various
friction coefficients is illustrated in Fig. \ref{fig:tm_phi.01}.
The characteristic feature of the temporal evolution of disc dynamo,
which is present in all cases except the marginal case of $\kappa=1,$
is the overshooting of the equilibrium state and the subsequent oscillations
of the rotation rate and the magnetic flux. These oscillations, which
decay unless $\kappa=\lambda=0$ and the dynamo is driven by a constant
torque, imply that deviation from the quasi-stationarity becomes significant
when the disc rotation rate approaches the dynamo threshold $\Omega_{c}=1.$
This is because the effective magnetic relaxation time being reciprocal
of the growth rate diverges as $\sim(\Omega_{c}-\Omega)^{-1}$ when
$\Omega$ approaches $\Omega_{c}.$ As a result, the quasi-stationarity
inevitably breaks down in the vicinity of the dynamo threshold regardless
of how slowly the rotation rate is ramped up. Deviation from the quasi-stationary
solution of Eq. (\ref{eq:Phi-ND}):
\begin{equation}
\Phi(\Omega)=\frac{\Phi_{0}}{\Omega_{c}-\Omega},\label{eq:Phi-ROT}
\end{equation}
becomes obvious when the magnetic flux is plotted against the rotation
rate as in Fig. \ref{fig:flx-omg}. This numerical data can be used
to assess the possibility of recovering $\Omega_{c}$ from the best
fit to the quasi-stationary solution. Limiting the fit to $\Omega\le\Omega_{c}/2$,
where the solution is relatively quasi-stationary, we recover the
values of $\Omega_{c}$ which slightly vary with $\kappa$ and lie
in the shaded strips Fig. \ref{fig:flx-omg}. As seen, $\Omega_{c}$
produced by the best fit is just slightly higher than the true value
$(\Omega_{c}=1)$ for $\Gamma=0.01$ whereas the difference is about
$20\%$ when $\Gamma=0.1.$ 

\begin{figure}
\begin{centering}
\includegraphics[width=0.5\columnwidth]{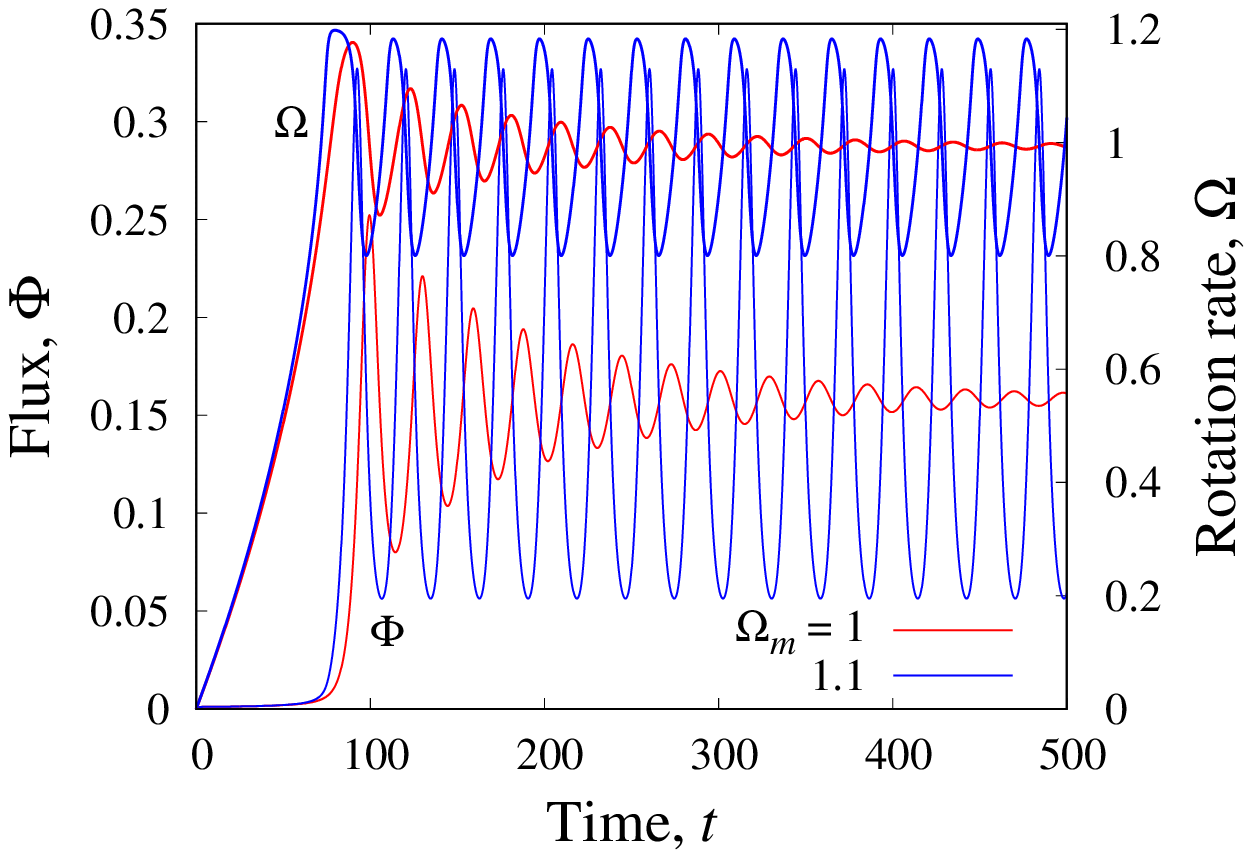}\put(-185,5){(a)}\includegraphics[width=0.5\columnwidth]{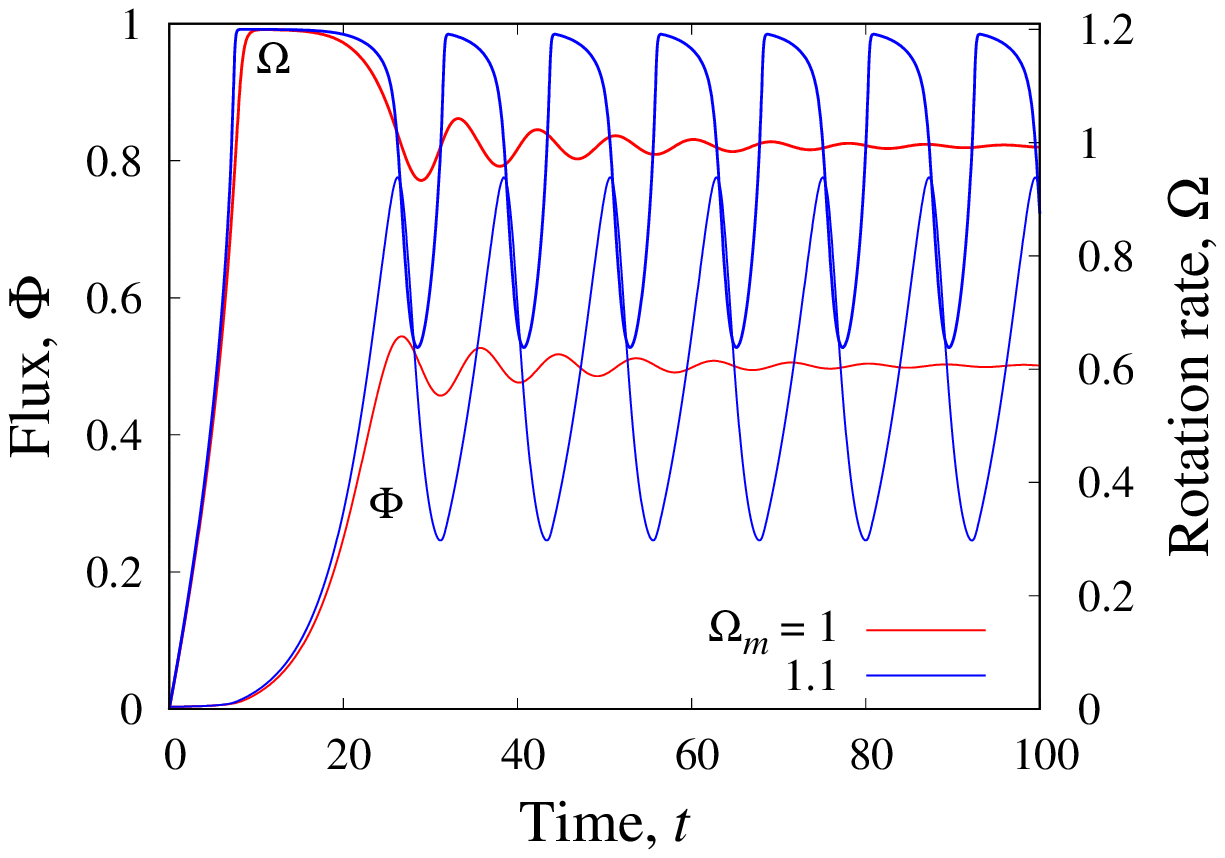}\put(-185,5){(b)}
\par\end{centering}
\caption{\label{fig:tm_phi.01_1.2}Temporal evolution of the rotation rate
$\Omega$ and the magnetic flux $\Phi$ in the presence of a background
field with $\Phi_{0}=0.01\Gamma_{\text{max}}^{1/2}$ and $\lambda=1$
when the disc is driven by an electric motor with synchronous frequency
$\Omega_{d}=1.2$ and the maximal torque $\Gamma_{\text{max}}=0.01\,\text{(a)},0.1\,\text{(b)}$
developed at $\Omega_{m}=1,\,1.1.$}
\end{figure}

When the dynamo is powered by an induction motor supplied with constant
frequency and voltage, the developed torque varies non-monotonously
with the rotation rate. The torque predicted by the basic asynchronous
electric motor model \citep{Hughes2019} can be written as
\begin{equation}
\Gamma=2\Gamma_{\text{max}}\frac{(\Omega_{d}-\Omega)(\Omega_{d}-\Omega_{m})}{(\Omega_{d}-\Omega)^{2}+(\Omega_{d}-\Omega_{m})^{2}},\label{eq:torque}
\end{equation}
where $\Omega_{d}$ and $\Omega$ are the synchronous and actual rotation
frequencies, $\Omega_{m}$ is the frequency at which the torque attains
its peak value $\Gamma_{\text{max}}.$ When $\Omega\approx\Omega_{d}$,
the torque varies approximately linearly with $\Omega_{d}-\Omega$:
\[
\Gamma\approx2\Gamma_{\text{max}}\frac{\Omega_{d}-\Omega}{\Omega_{d}-\Omega_{m}}.
\]
 When the frequency difference $\Omega_{d}-\Omega$ becomes sufficiently
large, the driving torque attains a maximum $\Gamma_{\text{max}}$
at the critical frequency $\Omega_{m}$ and then drops off asymptotically
as 
\[
\Gamma\sim2\Gamma_{\text{max}}\frac{\Omega_{d}-\Omega_{m}}{\Omega_{d}-\Omega}.
\]
 If the equilibrium rotation frequency $\Omega_{c}=1$ happens in
the latter operating regime, for small perturbations of the rotation
rate $\tilde{\Omega}=\Omega-1$, we have
\[
\Gamma\approx2\Gamma_{\text{max}}\frac{\Omega_{d}-\Omega_{m}}{\Omega_{d}-1}\left(1+\frac{\tilde{\Omega}}{\Omega_{d}-1}\right),
\]
where the second term is analogous to a friction force but with negative
coefficient 
\[
-2\Gamma_{\text{max}}\frac{\Omega_{d}-\Omega_{m}}{(\Omega_{d}-1)^{2}}.
\]
 When this negative friction outweighs the electromagnetic damping
with the coefficient (\ref{eq:ktld}), the stationary dynamo state
becomes unstable giving rise to periodic oscillations as in the original
Bullard dynamo model. This instability is illustrated in Fig. \ref{fig:tm_phi.01_1.2}
where the oscillations can be seen to cease decaying when the motor
is driven by a fixed frequency $\Omega_{d}=1.2$ and the frequency
at which the torque attains maximum is increased from $\Omega_{m}=1$
to $1.1.$

\section{\label{sec:discussion}Discussion}

The measured coil voltage allows us to assess the electric current
and the associated ohmic heating. Firstly, using the coil resistance
\citep{Priede2013}
\[
R_{c}=\frac{1+\beta^{2}}{2\pi d\sigma_{\text{Cu}}}\ln\frac{r_{o}}{r_{i}}\approx0\unit[.61]{\mu\Omega,}
\]
and the Ohm's law, we can estimate the current generated by the dynamo
at the coil voltage $V=\unit[10]{mV}$ as $I=V/R_{c}\approx\unit[16]{kA}.$
Then the associated power of ohmic heating in the coil and disc with
the resistivity $R_{d}=R_{c}(1+\beta^{2})^{-1}$ is 
\begin{equation}
P=I^{2}(R_{c}+R_{d})=\frac{V^{2}}{R_{c}}\left(1+(1+\beta^{2})^{-1}\right)\approx\unit[200]{W.}\label{eq:pwr}
\end{equation}
This power is significant relative to the maximal power the motor
is expected to produce when operating outside its optimal regime with
a large slip. Additional, but presumably insignificant, ohmic as well
as viscous power dissipation is expected in the liquid metal contacts.

\begin{figure}
\begin{centering}
\includegraphics[width=0.5\columnwidth]{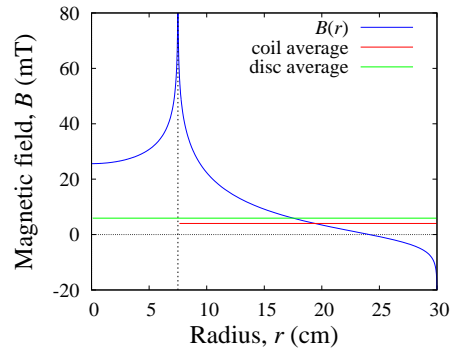}
\par\end{centering}
\caption{\label{fig:Brad}Radial distribution of the vertical magnetic field
$B(r)$ along the top surface of the coil computed using a quasi-stationary
current-sheet approximation \citep{Priede2013} for the coil voltage
$V=\unit[10]{mV}$ which corresponds to the dynamo current $I\approx\unit[16]{kA}.$
The two horizontal lines show the average values for the coil and
disc, respectively.}
\end{figure}

On the other hand, using the current estimate above, the average magnetic
flux density produced by the helical coil can be evaluated as
\begin{equation}
\bar{B}_{c}=\frac{\beta\text{\ensuremath{\mu_{0}}}Ir_{o}\bar{\phi}(\tilde{r}_{i}/r_{o})}{4\pi S}\approx\unit[4]{mT.}\label{eq:Bav}
\end{equation}
where $S=\pi(r_{o}^{2}-\tilde{r}_{i}^{2})$ is the area of the helical
part of the coil, $\beta=\tan(58^{\circ})\approx1.6$ is the helicity
of the logarithmic spiral arms, $\tilde{r}_{i}/r_{o}\approx1/4$ is
the ratio of the outer and inner radii of the coil and $\bar{\phi}(1/4)\approx1.38$
is the respective dimensionless flux \citep{Priede2013}. It is interesting
to note that this estimate is significantly lower than the magnetic
field strength measured on the surface of the coil. There may be several
reasons for this difference. Firstly, the distribution of the magnetic
field along the surface of the coil is highly non-uniform. In the
current-sheet approximation, the vertical flux density $B_{z}=r^{-1}\partial(rA_{\phi})$,
which is plotted in Fig. \ref{fig:Brad} using the analytical solution
for the vector potential $A_{\phi}$ obtained in \citep{Priede2013},
becomes unbounded at the inner and outer radii of the coil, where
it has logarithmic singularities. Besides that the magnetic flux density
can be seen in Fig. \ref{fig:Brad} to change direction close to the
outer edge of the coil. As a result, the average flux density is much
lower than the local field strength. Nevertheless, the field strength
at the centre of the coil, which represents a well-defined characteristic
value for this system, is comparable to the field strength measured
in the vicinity of the inner radius of the coil. Secondly, regardless
of the current-sheet approximation, there are logarithmic singularities
in the field strength at the sharp edges of the coil which may affect
the readings of the Hall sensor when placed directly atop the coil. 

Thirdly, and most importantly, the magnetic field distribution is
strongly affected by the iron frame of the set-up, especially by the
four crossed arms holding the coil. . As the Hall sensors were placed
on the coil next to these beams, they could be affected by the field
concentration caused by these ferromagnetic elements. The effect of
the iron frame is evident in the background magnetic field which was
found to be $\sim\unit[1]{mT,}$ \emph{i.e}., more than an order of
magnitude stronger than Earth's field. This fact implies that the
iron frame can have a similar magnitude effect also on the magnetic
field generated by the coil.

\begin{figure}
\begin{centering}
\includegraphics[width=0.5\columnwidth]{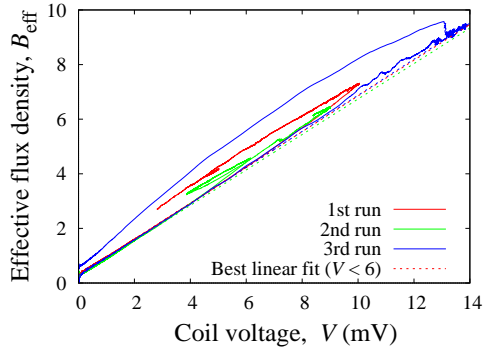}
\par\end{centering}
\caption{\label{fig:Beff}Effective magnetic flux density $B_{\text{eff}}$
(\ref{eq:Beff}) against the coil voltage $V$ for three runs along
with the best linear fit $B_{\text{eff}}=B_{0}+\kappa V$ for $V<\unit[6]{mV}$
which yields $B_{0}=\unit[0.29\pm0.03]{mT}.$}
\end{figure}

Given the relative smallness of the coil area covered by the frame,
the latter is expected to have a comparably small effect on the average
magnetic flux density. This is confirmed by the effective magnetic
flux density through the disc $B_{\text{eff}}$ which can be estimated
by comparing the e.m.f generated by the disc, $\Omega B_{\text{eff}}(r_{o}^{2}-r_{i}^{2})/2,$
and the total voltage drop over the disc and coil. The latter can
be estimated using the ohmic heating power $P$ (\ref{eq:pwr}) as
$P/I,$ which results in

\begin{equation}
B_{\text{eff}}=2V\frac{1+(1+\beta^{2})^{-1}}{\Omega(r_{o}^{2}-r_{i}^{2})}.\label{eq:Beff}
\end{equation}
As seen in Fig. \ref{fig:Beff}, $B_{\text{eff}}$ varies nearly linearly
with the coil voltage as predicted by the quasi-stationary approximation
well outside this regime. The best fit of this variation for $V<\unit[6]{mV}$
produces an effective flux density of the background magnetic field
$\bar{B}_{0}=\unit[0.29\pm0.03]{mT},$ which is about a factor of
three lower than the strength of the background field measured under
one of the four iron beams holding the coil.

On the other hand, the effective magnetic flux density, which reaches
$B_{\text{eff}}\approx\unit[7]{mT}$ at the coil voltage $V=\unit[10]{mV}$,
is somewhat higher than $\bar{B}_{c}$ defined by Eq. (\ref{eq:Bav}).
This difference may be due to the inner radius of the disc $r_{i}=\unit[4.5]{cm}$
being smaller than the respective radius of the helical part of the
coil $\tilde{r}_{i}=\unit[7.5]{cm}.$ As a result, the inner part
of the disc with $r<\tilde{r}_{i}$ is exposed to a relatively strong
magnetic field which is generated by the coil in this region (see
Fig. \ref{fig:Brad}). On the other hand, the respective magnetic
flux density averaged over the whole area of the disc, $\bar{B}_{d}\approx\unit[6]{mT,}$
is relatively close to $B_{\text{eff}}.$

\begin{figure}
\begin{centering}
\includegraphics[width=0.5\columnwidth]{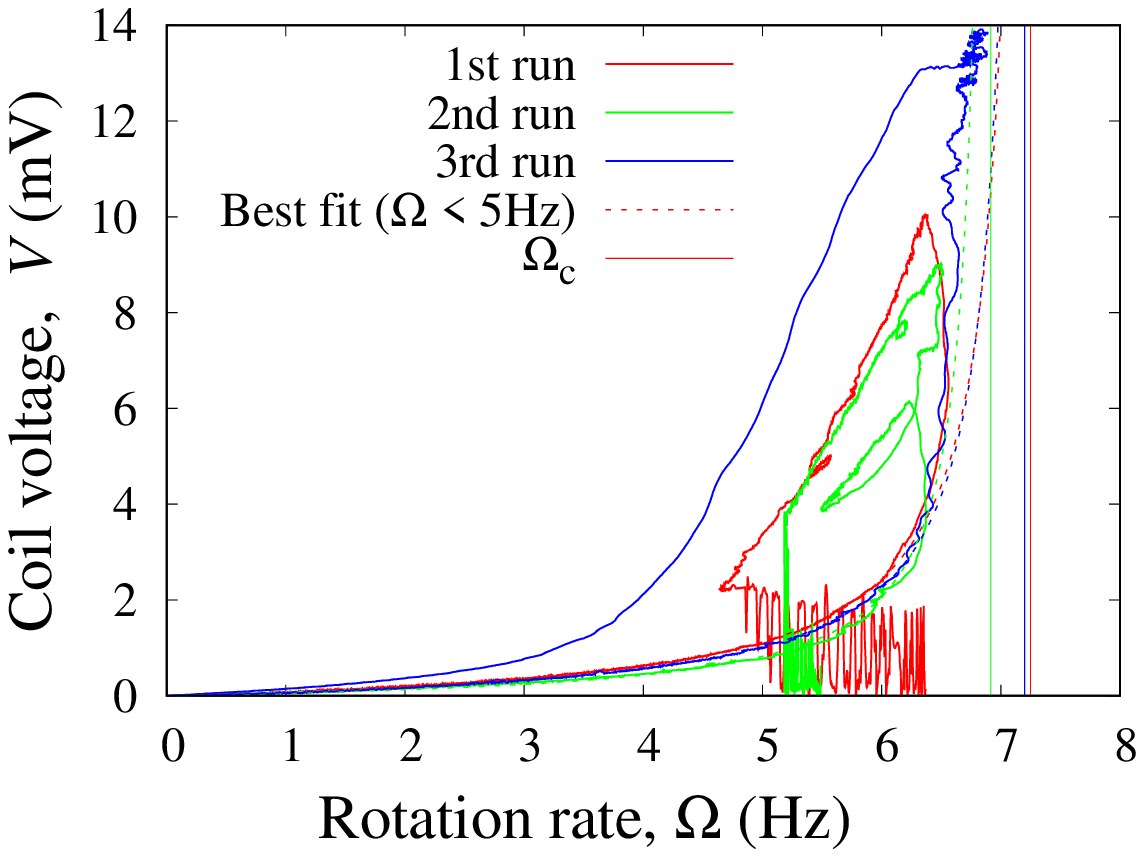}\put(-185,5){(a)}\includegraphics[width=0.5\columnwidth]{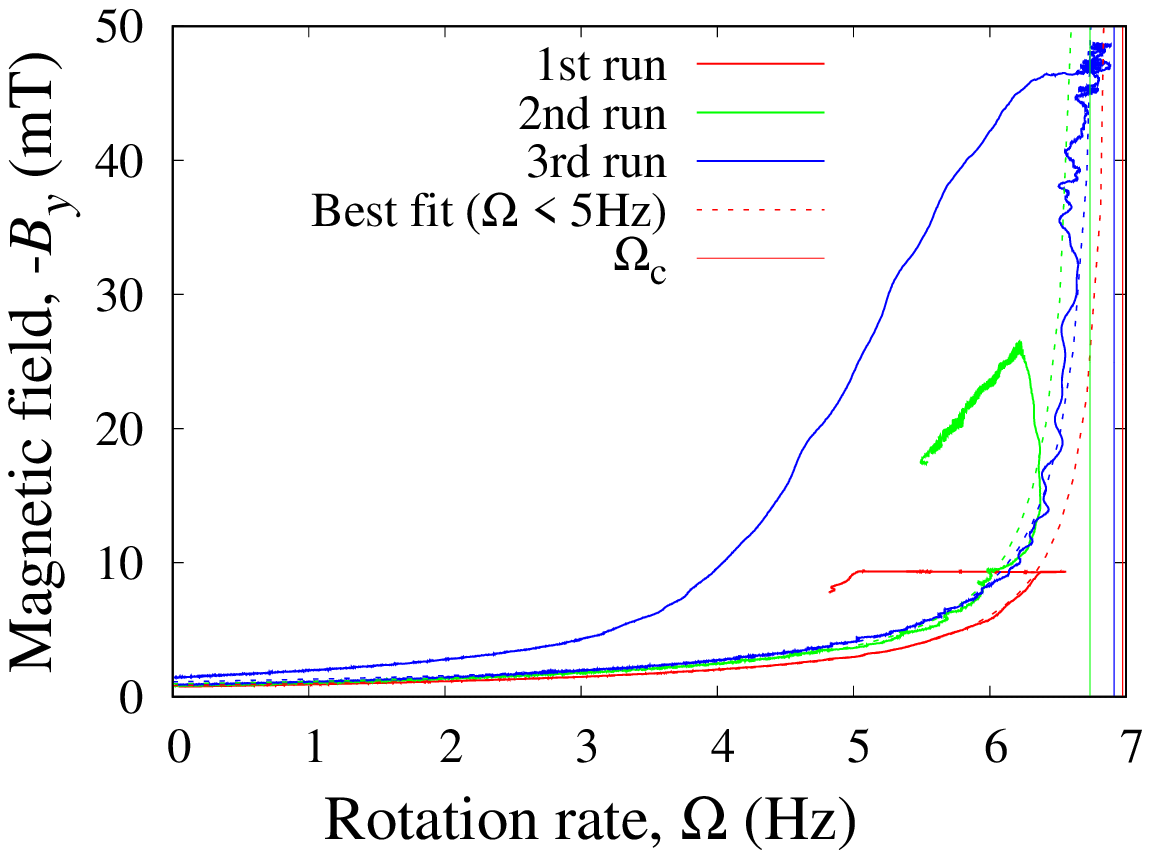}\put(-185,5){(b)}
\par\end{centering}
\caption{\label{fig:VB-ROT}The coil voltage (a) and the vertical component
of the magnetic field at the inner radius of the coil (b) versus the
rotation rate along with the best fits of the quasi-stationary solution
(\ref{eq:V-ROT},\ref{eq:B-ROT}) for $\Omega<\unit[5]{Hz.}$}
\end{figure}

The critical frequency $\Omega_{c}$ can be determined by fitting
the quasi-stationary solution (\ref{eq:Phi-ROT}), which can be written
in terms of the respective quantities as 
\begin{eqnarray}
V(\Omega) & = & \frac{V_{0}\Omega}{\Omega_{c}-\Omega},\label{eq:V-ROT}\\
B(\Omega) & = & \frac{B_{0}\Omega_{c}}{\Omega_{c}-\Omega},\label{eq:B-ROT}
\end{eqnarray}
to the variation of the coil voltage and magnetic field with the rotation
rate, which is plotted in Fig. \ref{fig:VB-ROT}. Limiting the fit
to $\Omega<\unit[6]{Hz},$ where the variation is expected to be sufficiently
close to quasi-stationary, and using the coil voltage, we find $\Omega_{c}\approx\unit[7.2,6.9,7.2]{Hz}$
for the respective run (see Fig. \ref{fig:VB-ROT}a). The best fit
of the magnetic field yields $\Omega_{c}\approx\unit[7.0,6.7,6.9]{Hz},$
which are somewhat lower than the respective previous values. The
last value is also slightly higher than the apparent equilibrium rotation
rate in the third run which may be seen in Fig. \ref{fig:tim-ser}
to reach $\approx\unit[6.8]{Hz}.$ 

Alternatively, $\Omega_{c}$ can be estimated using the the short
equilibrium state which appears in the third run between $t\approx\unit[60]{s}$
and $\unit[70]{s,}$ where $\Omega\approx\unit[6.8]{Hz},$$B\approx\unit[45]{mT}$
and $V\approx\unit[12.9]{mV.}$ Substituting these values into equations
(\ref{eq:V-ROT}) and (\ref{eq:B-ROT}) and using $B_{0}\approx\unit[1.1]{mT}$
and $V_{0}\approx\unit[0.45]{mT},$ which follow from the best of
$B$ and $V$ for this run, after a few rearrangements, we obtain
$\Omega_{c}=\Omega/(1-B_{0}/B)\approx\unit[6.97]{Hz}$ and $\Omega_{c}=\Omega(1+V_{0}/V)\approx\unit[7.0]{Hz}.$
As seen both results are practically identical and lying between the
values obtained from the best fit of quasi-stationary solutions for
$V$ and $B.$ Note that the theoretical dynamo threshold for our
set-up, which is defined for negligible contact resistance by the
minimal critical magnetic Reynolds number $\Rm\approx35$ \citep{Priede2013},
corresponds to $\Omega_{c}\approx\unit[8.2]{Hz}.$ Although the background
magnetic field makes the equilibrium rotation rate somewhat lower
than the self-excitation threshold, the difference between the theoretical
threshold and that extracted from the best fit of experimental results
may be due to the current sheet approximation used in the theoretical
model. 

The effective background magnetic field $B_{0}$ can also be estimated
using the best-fit parameters $V_{0}$ and $\Omega_{c}$ of the coil
voltage. Substituting $V$ from Eq. (\ref{eq:V-ROT}) into Eq. (\ref{eq:Beff})
and taking $\Omega\rightarrow0,$ it is easy see that $\bar{B}_{0}$
is obtained by replacing $V$ and $\Omega$ in Eq. (\ref{eq:Beff})
with $V_{0}$ and $\Omega_{c}.$ For the respective run, this yields
$\bar{B}_{0}\approx\unit[0.33,0.26,0.30]{mT,}$ which are consistent
with the previous estimate obtained from the best fit of the effective
magnetic flux variation with the coil voltage plotted in Fig. \ref{fig:Beff}.
The larger scatter in the last estimate is because it is obtained
by using low rotation rates only, whereas the previous relation holds
in a much larger range of $\Omega$ which may be seen in Fig. \ref{fig:Beff}
to extend well beyond the quasi-stationary regime.

\section{\label{sec:conclusion}Summary and conclusions}

In this paper, we presented experimental results from three successful
runs of a Bullard-type homopolar disc dynamo. The realisation of such
a dynamo was commonly thought impossible because of the prohibitively
high rotation rates which are required when sliding electric contacts
are made of graphite brushes. We overcome this problem by using $\mathit{GaInSn}$
eutectic alloy, which is liquid at room temperature, for sliding electric
contacts. The set-up consisted of a copper disc with a radius of $\unit[30]{cm}$
and thickness of $\unit[3]{cm}$ which was placed co-axially beneath
a flat, multi-arm spiral coil of the same size and connected to it
electrically at the centre and along the perimeter by $\mathit{GaInSn}$
contacts \citep{Priede2013}. The dynamo was effectively axisymmetric
but anisotropic because of the spiral slits which deflected the current
in the coil azimuthally so generating an axial magnetic field. 

The use of liquid metal in sliding electrical contacts came with two
complications. Firstly, the liquid metal was expelled from the peripheral
contact by centrifugal force radially inwards over the top surface
of the coil. This problem was largely mitigated by reducing the annular
gap between the disc and the coil from $\unit[3]{mm}$ in the original
design \citep{Avalos-Zuniga2017} to $\unit[0.25]{mm}$ in this set-up.
Secondly, as the liquid metal was exposed to air, it quickly oxidized
when the device ran. This limited the time of the experiment to a
few minutes after which the electric contact between the disc and
the coil usually failed. 

The runs differed mainly by how the motor driving frequency $\Omega_{d}$
was varied. In the first run, $\Omega_{d}$ was increased at a nearly
constant rate and the magnetic field was measured using only the low-field
probe which was placed on the coil in its central part close to the
inner radius of the spiral slits. When the disc rotation rate reached
$\Omega\approx\unit[6.5]{Hz},$ the vertical field component was found
to exceed $\unit[9]{mT},$ which was the upper detection limit of
this probe. Although the driving frequency was set at $\Omega_{d}=\unit[7.33]{Hz},$
the rotation rate started to fall after reaching $\approx\unit[6.7]{Hz.}$
It means that the electromagnetic torque braking the disc had exceeded
the breakdown torque of the electric motor, which, thus, started to
stall. Extrapolation using the voltage drop across the coil, which
was measured in addition to the magnetic field, indicated that the
vertical field strength in the first run had reached $\approx$$\unit[25]{mT}.$
It is important to note that the magnetic field in the vicinity of
the inner radius of the coil may be significantly higher than the
average over the coil surface.

In the second run, the low-field probe was moved to the outer radius
and the medium-field probe was installed in its place. Besides that,
the driving frequency was ramped up through intermediate constant
steps rather than continuously as in the first run. We again observed
a steep increase of the magnetic field with $B_{y}$ at the inner
radius reaching $\unit[26.5]{mT}$ when the rotation rate approached
$\unit[6.4]{Hz}.$ At this point, the rotation rate started to fall
as in the first run. However, the second run was sufficiently long
for the disc to start spinning up again. The re-acceleration was enhanced
by ramping up $\Omega_{d}$ to $\unit[8.47]{Hz}.$ This resulted in
$\Omega$ reaching $\approx\unit[6.5]{Hz}$ and $B_{y}$ shooting
up to $\unit[\approx38]{mT.}$ At this point, the motor started to
stall again and the experiment was terminated because of the loss
of electric contact. 

Stalling was not observed in the third run, in which the driving frequency
$\Omega_{d}$ was set higher and initially ramped up faster. With
$\Omega_{d}\approx\unit[8.9]{Hz}$ reached in $\unit[\approx60]{s},$
the rotation rate saturated at $\Omega\approx\unit[6.6\pm0.1]{Hz}$
after $\unit[\approx45]{s}.$ The magnetic field kept growing as long
as $\Omega_{d}$ increased saturating at $\unit[\approx45]{mT.}$
After $\unit[\approx30]{s,}$ when $\Omega_{d}$ was raised to $\approx\unit[9.7]{Hz},$
$B_{y}$ increased to $\unit[\approx48.6]{mT}$ without a noticeable
change in the rotation rate $\Omega.$ This is a characteristic behaviour
of a fully developed disc dynamo. 

We also proposed an extended disc dynamo model which qualitatively
reproduces experimental results by taking into account the background
magnetic field, transient eddy currents in the disc as well as the
non-linearity of the electric motor. The background magnetic field,
which was found to be an order of magnitude stronger than Earth's
magnetic field, was obviously due to the iron frame of the set-up.
At sub-critical rotation rates, i.e., those below the kinematic threshold,
the dynamo works as a homopolar generator amplifying the background
magnetic field. In quasi-stationary approximation, the amplification
rate increases with the rotation frequency and tends to infinity at
the dynamo threshold. We used this fact to determine the dynamo threshold
from the best fit of the magnetic field and the coil voltage. In this
way, we found a critical rotation frequency $\Omega_{c}\approx\unit[7\pm0.2]{Hz},$
which is somewhat larger than the actual saturation frequency observed
in the third run. This is consistent with the numerical results which
showed that the transient effects result in the best-fit value being
somewhat higher than the actual dynamo threshold. On the other hand,
this experimental result is somewhat lower than the theoretical prediction
$\Omega_{c}\approx\unit[8.2]{Hz}$ corresponding to the minimal critical
magnetic Reynolds number $\Rm\approx35$ for a negligible contact
resistance \citep{Priede2013}. This difference is likely due to the
very approximate nature of the current sheet model which was used
to evaluate the dynamo threshold. In particular, due to the strongly
uniform radial magnetic flux distribution (see Fig. \ref{fig:Brad}),
the total magnetic flux through the disc may be larger than theoretically
predicted because the effective inner radius of the disc is smaller
than that of the coil. The critical rotation rate can be reduced further
by optimizing the inner and outer disc radii which were originally
chosen to be the same as those of the coil.

\ack{We are grateful to Adrian Pérez, Carla Bello and and Manuel Valencia for the technical support.} 
\funding{This work was supported by the National Council of Science and Technology of Mexico (CONACYT) through grant CB-168850 and the National Polytechnic Institute (IPN) through grants SIP-20211736 and SIP-20220849.} 

\bibliographystyle{RS}
\bibliography{/home/priede/work/dynamo/ref/dynamo}

\end{document}